\documentclass[11pt,a4paper]{article}
\pdfoutput=1

\usepackage{jcappub, natbib}
\usepackage{ifthen}
\usepackage{graphicx}% Include figure files
\usepackage{dcolumn}% Align table columns on decimal point
\usepackage{bm}% bold math
\usepackage{color}
\usepackage{amsmath}
\usepackage{amssymb}
\usepackage{bbold}
\usepackage{longtable}

\newcommand{\rperp}{r_\perp}
\newcommand{\rpar}{r_\parallel}
\newcommand{\xioned}{\xi_{\rm 1D}}
\newcommand{\pkoned}{P_{\rm 1D}}

\newcommand{\cm}{\ \text{cm}}

\newcommand{\Mpc}{\ \text{Mpc}}
\newcommand{\hMpc}{\ h^{-1}\text{Mpc}}

\newcommand{\iMpc}{\ \text{Mpc}^{-1}}

\newcommand{\be}{\begin{equation}}
\newcommand{\ee}{\end{equation}}

\renewcommand{\vec}{\bm}

\newcommand{\kms}{{\rm km~s^{-1}}}

\newcommand{\lya}{Ly$\alpha$}

\newcommand{\lyaf}{Ly$\alpha$ forest}

\begin{document}

\title{Mock Quasar-Lyman-$\alpha$ Forest Data-sets for the
SDSS-III Baryon Oscillation Spectroscopic Survey}
\author[a,b]{Julian E. Bautista,}
\author[c]{Stephen Bailey,}
\author[c]{Andreu Font-Ribera,}
\author[d,e]{Matthew M. Pieri,}
\author[a,f,g]{Nicolas G. Busca,}
\author[h,i]{Jordi Miralda-Escud\'e,}
\author[j]{Nathalie Palanque-Delabrouille,}
\author[j,a]{James Rich,}
\author[b]{Kyle Dawson,}
\author[k]{Yu Feng,}
\author[l]{Jian Ge,}
\author[h]{Satya Gontcho A Gontcho,}
\author[k]{Shirley Ho,}
\author[j]{Jean Marc Le Goff,}
\author[m]{Pasquier Noterdaeme,}
\author[m]{Isabelle P\^aris,}
\author[n,f]{Graziano Rossi,}
\author[c]{David Schlegel}

\affiliation[a]{APC, Universit\'{e} Paris Diderot-Paris 7, CNRS/IN2P3, CEA, 
	Observatoire de Paris, 10, rue A. Domon \& L. Duquet,  Paris, France}
\affiliation[b]{Department of Physics and Astronomy, University of Utah, 115 S 100 E, RM 201, Salt Lake City, UT 84112, USA}
\affiliation[c]{Lawrence Berkeley National Laboratory,
	1 Cyclotron Road, Berkeley, CA, USA}
\affiliation[d]{Aix Marseille Universit\'e, CNRS, LAM (Laboratoire d'Astrophysique de Marseille) UMR 7326, 38 rue Frédéric Joliot-Curie,  13388, Marseille, France}
\affiliation[e]{Institute of Cosmology \& Gravitation, University of Portsmouth, 1-8 Burnaby Road, Portsmouth PO1 3FX, UK}
\affiliation[f]{Observat\'orio Nacional, Rua Gal. Jos\'e Cristino 77, Rio de Janeiro, RJ - 20921-400, Brazil}
\affiliation[g]{25 Laborat\'orio Interinstitucional de e-Astronomia, - LIneA, Rua Gal.Jos\'e Cristino 77, Rio de Janeiro, RJ - 20921-400, Brazil}
\affiliation[h]{Institut de Ci\`encies del Cosmos, Universitat de Barcelona/IEEC, 1 Mart\'i i Franqu\`es, Barcelona 08028, Catalonia, Spain}
\affiliation[i]{Instituci\'{o} Catalana de Recerca i Estudis  Avan\c{c}ats, 23 Passeig Llu\'is Companys, Barcelona, Catalonia}
\affiliation[j]{CEA, Centre de Saclay, Irfu/SPP, D128, F-91191 Gif-sur-Yvette, France}
\affiliation[k]{McWilliams Center for Cosmology, Carnegie Mellon University, 5000 Forbes Avenue, Pittsburgh, PA, 15213, USA}
\affiliation[l]{Department of Astronomy, University of Florida, 211 Bryant Space Science Center, Gainesville, FL 32611-2055, USA}
\affiliation[m]{Universit\'e Paris 6 et CNRS, Institut d'Astrophysique de Paris, 98bis blvd. Arago, 75014 Paris, France}
\affiliation[n]{Department of Astronomy and Space Science, Sejong University, 209 Neungdong-ro, Gwangjin-gu, Seoul, 143-747, Korea}

\abstract{
We describe mock data-sets generated to simulate the 
high-redshift quasar sample in 
Data Release~11 (DR11) of the SDSS-III Baryon Oscillation Spectroscopic 
Survey (BOSS).
The mock spectra contain \lyaf\ correlations useful for studying the 
3D correlation function including Baryon Acoustic Oscillations (BAO).
They also include astrophysical effects such as 
quasar continuum diversity and 
high-density absorbers,
instrumental effects such as noise and spectral resolution, 
as well as 
imperfections introduced by the SDSS pipeline treatment of the raw data.
The \lyaf\ BAO analysis of the BOSS collaboration, described in Delubac et al. 2014,
has used these mock data-sets to 
develop and cross-check analysis procedures 
prior to performing the BAO analysis
on real data, and for continued systematic cross checks.
Tests presented here show that
the simulations reproduce sufficiently well 
important characteristics of real spectra.
These mock data-sets will be made available together with the data at the time
of the Data Release~11.
}


\maketitle

\section{Introduction}
\label{sec:intro}

The Baryon Oscillation Spectroscopic Survey (BOSS) \cite{2013AJ....145...10D}, 
as part of the Sloan Digital Sky Survey \cite{2011AJ....142...72E},
has obtained accurate redshifts of over 1.5 million massive galaxies 
and spectra of more than 150,000 quasars with \lyaf\ coverage in order 
to study Baryon Acoustic Oscillations (BAO). 
Measuring large-scale structure with 
these tracers provides percent-level accuracy in the position of the 
BAO peak \cite{
2014arXiv1411.1074A}, 
which translates into measurements of the redshift dependent
angular diameter distance, $D_A(z)$, and  Hubble expansion rate, $H(z)$.

%\af{I would also add a reference to the joint paper here, may be instead of 
%Weinberg? After all we are talking about BAO with BOSS. May be both. }
 
BOSS Data Release 9 (DR9) was used to study BAO
at high redshift for the first time 
in the 3D \lya\ forest flux distribution 
\citep{2011JCAP...09..001S, 2013A&A...552A..96B, 2013JCAP...04..026S,2013JCAP...03..024K}. 
% This sample contained $\sim 50,000$ \lya\ forests, 
%and obtained a 3\% measurement of the expansion rate at $z \sim 2.3$.
%\jr{This needs revision in light of DR11 results:}
%Combined with the latest BAO measurement from BOSS CMASS galaxies 
%\cite{2013arXiv1312.4877A}, the \lya\ BAO results confirms that
%the universe was still decelerating  at this redshift, as expected in the 
%concordance $\Lambda$CDM model.
%
The latest \lya\ forest clustering analysis, 
described in \cite{delubac_baryon_2015}, 
uses three times more data from 
Data Release 11 (DR11\footnote{Publicly available
in December 2014 along with our mock catalogs.}).
%The Data Release 11, data and mock catalogs, will be publicly available in 
%December 2014 together with DR12.}). 
This new analysis is an improvement in several ways: new covariance matrix 
estimates, 
deeper study of systematic effects and the use of 100 realizations of
mock catalogs. It has yielded a new measurement of the expansion rate, 
$H(z=2.34) = (222\pm7) (147.4/r_d) ~\kms\iMpc$, and the angular distance 
$D_A(z) = (1662\pm96) (r_d/147.4) ~\Mpc$, where $r_d$ is the sound horizon at 
the drag epoch 
as can be estimated by CMB data \cite{2013arXiv1303.5076P}.

Mock catalogs were essential in these previous analyses
for testing systematic effects in the BAO measurements. 
We describe for the first time in this work 
the generation and properties of 
mock spectra used in the DR11 \lya\  
analysis. The DR9 mock catalogs used the same procedure described here, 
the only differences being the size of 
the quasar sample (33\% of DR11) and the number of available realizations 
(15 instead of 100 in DR11). 
The larger number of realizations were
important to increase the precision in the determination of
possible biases in the analysis, for instance in the 
estimate of the covariance matrix of our measurements,
the effect of the absorption by metals in the intergalactic medium,
and the effect of errors in the spectrophotometric reduction.

The clustering of the \lya\ forest is encoded in the 
absorption field defined by
\begin{equation}
\delta_q(\lambda)=\frac{F_q(\lambda)}{\bar{F}(\lambda)}-1
\hspace*{15mm}
F_q(\lambda)= \frac{f_q(\lambda)}{ C_q(\lambda)}
\label{eq:deltafield}
\end{equation}
where $q$ indexes an individual quasar (or more precisely its
angular position on the sky) and $\lambda$ is 
the observed-frame wavelength of the \lya\ absorption. 
The flux transmission, $F_q(\lambda)$,
and its mean $\bar{F}(\lambda)$, is
mostly due to neutral hydrogen at redshift $z=\lambda/(121.6$~nm$)-1$.
The transmission is
the ratio of the measured flux, $f_q(\lambda)$,  to
the quasar ``continuum'' $C_q(\lambda)$, i.e. the
flux that would be measured in the absence of absorption. 
The $\delta_q(\lambda)$ can be used to estimate the two-point correlation 
function
\begin{equation}
\hat{\xi}_A = \frac{\sum_{(i,j)\in A}w_{ij}\delta_i\delta_j}{\sum_{(i,j)\in A}w_{ij}}
\label{eq:2ptcorrelation}
\end{equation}
where $i$ and $j$ refer to pixels defined by $q$ and $\lambda$ and
the sum is over pixel pairs $(i,j)$ such that the separation between
the two pixels lies in bin $A$ (generally defined by a transverse separation,
$\rperp$ and radial separation $\rpar$).
The $w_{ij}$ are weights that depend on the level of instrumental noise in the pixels and on the intrinsic variance of fluctuations in the forest.

The creation of realistic mocks requires generating
realistic quasar continua, mean absorptions and $\delta_q(\lambda)$ that
have correlations due to large-scale structure and fluctuations due to noise.
This is done in two steps.
First, we create 
a realization of the absorption field at every forest pixel, 
assuming a certain cosmological model. 
Second, quasar spectra are generated by multiplying the $F_q(\lambda)$
by synthetic quasar continua, $C_q(\lambda)$, and then adding 
instrumental noise, metal lines, 
high column density absorbers, and other potential systematics. 

Full hydrodynamical simulations \cite{1994ApJ...437L...9C, 1995A&A...295L...9P,  1996ApJ...471..582M, 2003ApJ...585...34M, 2014arXiv1401.6464R} can give a precise description of the 
\lya\ forest physics under certain assumptions, but they are computationally 
expensive and do not have the required dynamic range to describe both BAO 
scales and the small-scale structure 
responsible for the observed \lya\ fluctuations. 
Furthermore, hundreds of realizations of the survey are desired, 
which is impractical with these simulations. 
With the focus on BAO scales, 
an alternative method to generate a correlated absorption field was presented in
\cite{2012JCAP...01..001F}. 
This method is fast enough to allow us to generate many realizations of the
survey, while at the same time capturing the correct 1-point and 2-point 
statistics of the absorption field. This method has been used in previous analyses
\citep{2011JCAP...09..001S,2013A&A...552A..96B, 2013JCAP...04..026S},
and we also follow it here. 

The second step, the main subject of this article, is the transformation of the 
flux transmission into data-like spectra that have the
essential characteristics of the real BOSS spectra. 
Absorption lines arising from high column density 
systems and metal transitions are first added to the \lya\ absorption.
The absorption field is then applied to  random 
quasar continua and instrumental noise is added.  
Pipeline imperfections in flux and noise estimation, and sky subtraction
are included.

Throughout this work, 
%in order to transform between physical distances and
%observed angles and wavelengths,
we use a fiducial flat $\Lambda$CDM cosmology with  
$\Omega_m = 0.27$, $\Omega_bh^2 = 0.0227$, $h=0.7$, $\sigma_8=0.8$, 
$n_s=0.97$, $\Omega_\nu=0$.
This cosmology was also
used in previous papers on the \lya\ BAO measurements of BOSS
(\cite{2013A&A...552A..96B}, \cite{2013JCAP...04..026S},
\cite{2013JCAP...03..024K},\cite{2013arXiv1311.1767F}). 
Note, however, that the latest analysis presented in \cite{delubac_baryon_2015}
uses a fiducial model with massive neutrinos ($\Omega_\nu h^2=0.0006$).

We start by summarizing the method to generate the \lya\ absorption field
in section \ref{sec:absorption}. 
In section \ref{sec:expanded} we describe the transformation
this field, the so-called \emph{raw} mock spectra, into realistic
quasar spectra, the so-called \emph{expanded} mock spectra. 
In section \ref{sec:tests} we compare some basic statistics of the 
expanded spectra with those of real spectra.
In section \ref{sec:covsec} we compare the three-dimensional
correlation function and its covariance matrix of the expanded mocks
with those of the data.
We conclude in section \ref{sec:conc}. 
In appendix \ref{app:access} we explain the access and the usage of the mock catalogs. These can be found at the SDSS public webpage \url{http://www.sdss.org/dr12/algorithms/lyman-alpha-mocks}.

\section{Generation of the absorption field}
\label{sec:absorption}

%\af{We should decide whether to use $P_F$ or $P_{input}$ and stick to it
%within the section. I'd prefer to use $P_F$.}

The main desired properties of the absorption field are that 
(a) it possesses pre-defined 3D correlations described by a 
power spectrum $P_F(k)$,
(b) it has a realistic flux probability distribution function, with transmissions values between 0 and 1
and 
(c) it follows the geometry of the BOSS \lya\ survey, \emph{i.e.}, 
it contains quasars at the same position and redshift as
the real quasars, and has a wavelength resolution better than the BOSS spectrograph pixel size ($\sim 0.07$~nm).
We note that property (b) implies that the statistics of the absorption field will be non-Gaussian. 

We warn the user that these mocks are not suited for quasar-\lyaf\ cross-correlation measurements (e.g. \cite{2013arXiv1311.1767F}), since no correlations between quasar positions and the absorption field are included. 

\subsection{Input power spectrum}

The target correlation function of the absorption field is defined by the
power spectrum
\begin{equation}
P_F (k, \mu) = P_0^2 (1+ \beta \mu^2)^2 P_{\rm lin}(k) D(k, \mu)
\label{eq:power_spectrum}
\end{equation}
where $P_{\rm lin}(k)$ is the linear isotropic matter power spectrum as computed using CAMB\footnote{\url{http://camb.info/}} code \cite{2000ApJ...538..473L} at $z=2.25$ with cosmological parameters 
given in section \ref{sec:intro}. The amplitude $P_0=-0.14$ and redshift-distortion 
parameter $\beta=1.4$ were chosen to be compatible with measurements of 
\cite{2006ApJS..163...80M} and \cite{2011JCAP...09..001S}. 
The $D(k,\mu)$ term accounts for a small scale non-linear correction of the power
 spectrum (defined by Eq.~21 and Table~1 of \cite{2003ApJ...585...34M}), depending on the cosine of the angle 
 between the wavevector and the line of sight, $\mu = k_\parallel/k$. 

The redshift evolution of the power spectrum amplitude $P_0(z)$ is set by the evolution
of the \lyaf\ bias $b(z)$ and the growth factor $g(z)$, and we further assume
that it evolves according to
$b(z)\frac{g(z)}{g(z=2.25)}=-0.14[(1+z)/3.25]^{\gamma/2}$, where $\gamma = 3.8$ (following results from \cite{2006ApJS..163...80M}). The redshift space distortions parameter $\beta$ is assumed to be redshift-independent.

\subsection{Absorption field properties}

We use same assumption as in \cite{2012JCAP...01..001F}. In order to create an absorption field with a realistic flux probability density function, we build a Gaussian random field $\delta_G$ over pixels, and then we apply the following non-linear transformation to compute the transmission $F_q$,
\begin{equation}
F_q(\delta_G) = \exp\left[-a(z)e^{b(z)\delta_G}\right]\;.
\label{eq:transmission}
\end{equation}
The two functions, $a(z)$ and $b(z)$ are chosen so
that the resulting
mean transmission $\bar{F}$ and variance of $\delta_q = F_q/\bar{F}-1$ are
\begin{equation}
\ln\bar{F}(z) = \ln(0.8)
\left[ \frac{1+z}{3.25} \right]^{3.2}
\label{eq:meanabsorption}
\end{equation} 
\begin{equation}
{\rm Var}(\delta_q) = \frac{\sigma^2_F}{\bar{F}^2}(z)=0.108 \left[ \frac{1+z}{3.25} \right]^\gamma
\label{eq:absorptionvariance}
\end{equation}
where $\gamma=3.8$ is set by the evolution of the power spectrum amplitude.

Given this transformation between Gaussian variables and transmission, we proceed as follows: first, we generate a Gaussian field with correlations given by a Gaussian power spectrum $P_G$. Second, we compute transmissions using Eq.~\ref{eq:transmission}. By choosing the correct $P_G$, the final transmission field should have the correct correlations given by $P_F$ (Eq.~\ref{eq:power_spectrum}). 
The authors of \cite{2012JCAP...01..001F} give a prescription for choosing the 
power spectrum of the Gaussian field and the parameters of the non-linear 
transformation in order to match the desired correlation properties of 
the final absorption field. 

\subsection{The mock-data sample}

Our data sample is defined by the automatic object classification and redshift
determination performed by the BOSS reduction pipeline 
\citep{2012AJ....144..144B} and based on the BOSS Data Release 11 (DR11).
We selected all objects classified as a quasar and with redshifts in the range 
$2.15<z<3.5$, which captures the selection done for the analysis of real data 
\cite{delubac_baryon_2015}. 
This sample contains 149,751 line-of-sights and differs from the 
137,562-quasar sample in \cite{delubac_baryon_2015}, based on the DR11 quasar 
value-added catalog, mainly due to the rejection of quasars with broad
absorption lines (BALs) applied to the latter sample and, to a lesser
extent, to stars, galaxies or low-redshift quasars mis-identified as 
high-redshift quasars by the pipeline. 
This is not a problem since these extra spectra can be rejected at the 
analysis step. 
On the other hand, 6,808 quasars on the \cite{delubac_baryon_2015} data sample do 
not have a corresponding mock quasar.

Given the size of the BOSS \lyaf\ survey of over $10^5$ quasars, these conditions 
imply generating over $10^7$ correlated pixels. 
A straight-forward algorithm of generating uncorrelated pixels and taking 
linear combinations of those pixels to give them the desired correlations 
would involve inverting matrices of dimensions
$10^7\times10^7$, beyond the reach of current computational capabilities.

Fortunately, these difficulties are overcome by a method 
developed in \cite{2012JCAP...01..001F} that makes use
 of the fact that within a small angular region, 
quasar lines-of-sight are nearly parallel. 
We summarize this method here.
The desired correlated Gaussian random field,
$\delta_G(x_\parallel,\vec{x}_\perp)$, is defined as a function
of a radial coordinate, $x_\parallel$, and a transverse two-dimensional
coordinate vector, $\vec{x}_\perp$. This field is 
expressed in terms of its radial Fourier transform,
$\tilde{\delta}_G(k_\parallel,\vec{x}_\perp)$.
In the approximation of parallel lines-of-sight, 
the $\tilde{\delta}_G$ with differing $k_\parallel$
are uncorrelated and the statistical properties of $\tilde{\delta}_G$
of a given $k_\parallel$
are defined by the power spectrum
\begin{equation}
P_\times(k_\parallel,\vec{x}_\perp)= \frac{1}{2\pi}\int_{k_\parallel}^\infty
k_\perp {\rm d}k_\perp \frac{\sin (k_\perp r_\perp)}{k_\perp r_\perp}
P_G(k_\parallel,k_\perp) ~,
\end{equation}
where $P_G(k_\parallel,k_\perp)$ is the power spectrum of the Gaussian field
(not to be confused with $P_F$, the desired power for the final flux
field).

Correlated values of 
$\tilde{\delta}_G(k_\parallel,\vec{x}_\perp)$ at the angular
position of each quasar can thus be 
generated for each $k_\parallel$ by inverting a 
$N_{\rm qso}\times N_{\rm qso}$ matrix.
This is still beyond our computational capacity so
we divided the 
BOSS survey into regions of nearly 8,500 quasars each. The absorption 
field in each region is generated independently, as further detailed
below. Once the amplitudes and phases of the radial modes are fixed,
the Gaussian field is reconstructed by Fourier transforming the radial
modes for each line of sight.%}

\subsection{Implementation}

To ease the implementation of the radial Fourier transforms, the 
absorption fields so produced are sampled in fixed grid of
8192 bins of comoving width $0.5~h^{-1}\mathrm{Mpc}$ along each line
of sight, which covers a comoving distance of $4096 \hMpc$. 
They are then trimmed between the limits of the BOSS 
spectrograph ($\lambda \sim 360$~nm corresponding to $z \sim 1.96$) 
and the \lya\ emission peak of each quasar. We called this trimmed 
absorption field the ``raw mock skewers'', which is one of our products.
In a later step they are resampled at the center of BOSS wavelength pixels.

The redshift evolution of the absorption field is implemented by following 
this procedure at four redshift snap-shots: 1.920, 2.409, 2.898, and 3.386 
which cover the range of the BOSS \lya\ forests. 
In each snap-shot, the underlying density field is generated with the same 
seeds. 
This ensures that the final absorption field will look like an evolved field 
from snap-shot to snap-shot. 
We also shifted the positions of the lines-of-sight at each redshift accounting
for the fact that they are not parallel. 
The absorption field at any redshift is obtained by interpolation.

A total of 100 independent realizations were generated using the computational
facilities of NERSC \footnote{http://www.nersc.gov}.

\subsection{Effects of sky subdivisions}

\begin{figure}[t]
 \begin{center}
  \includegraphics[width=0.51\textwidth]{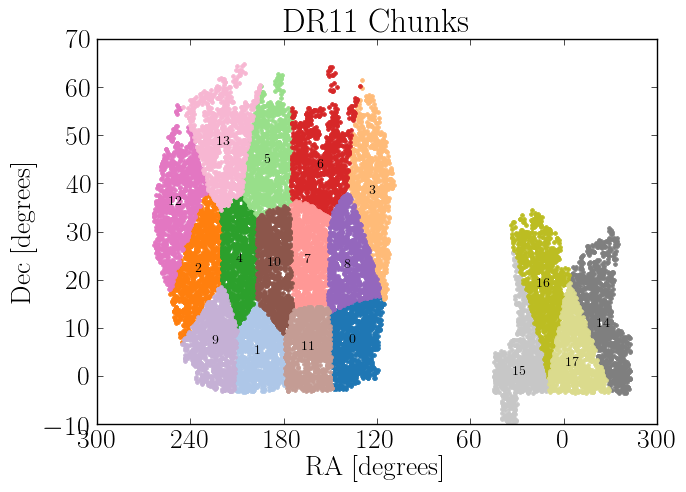}
   \includegraphics[width=0.46\textwidth]{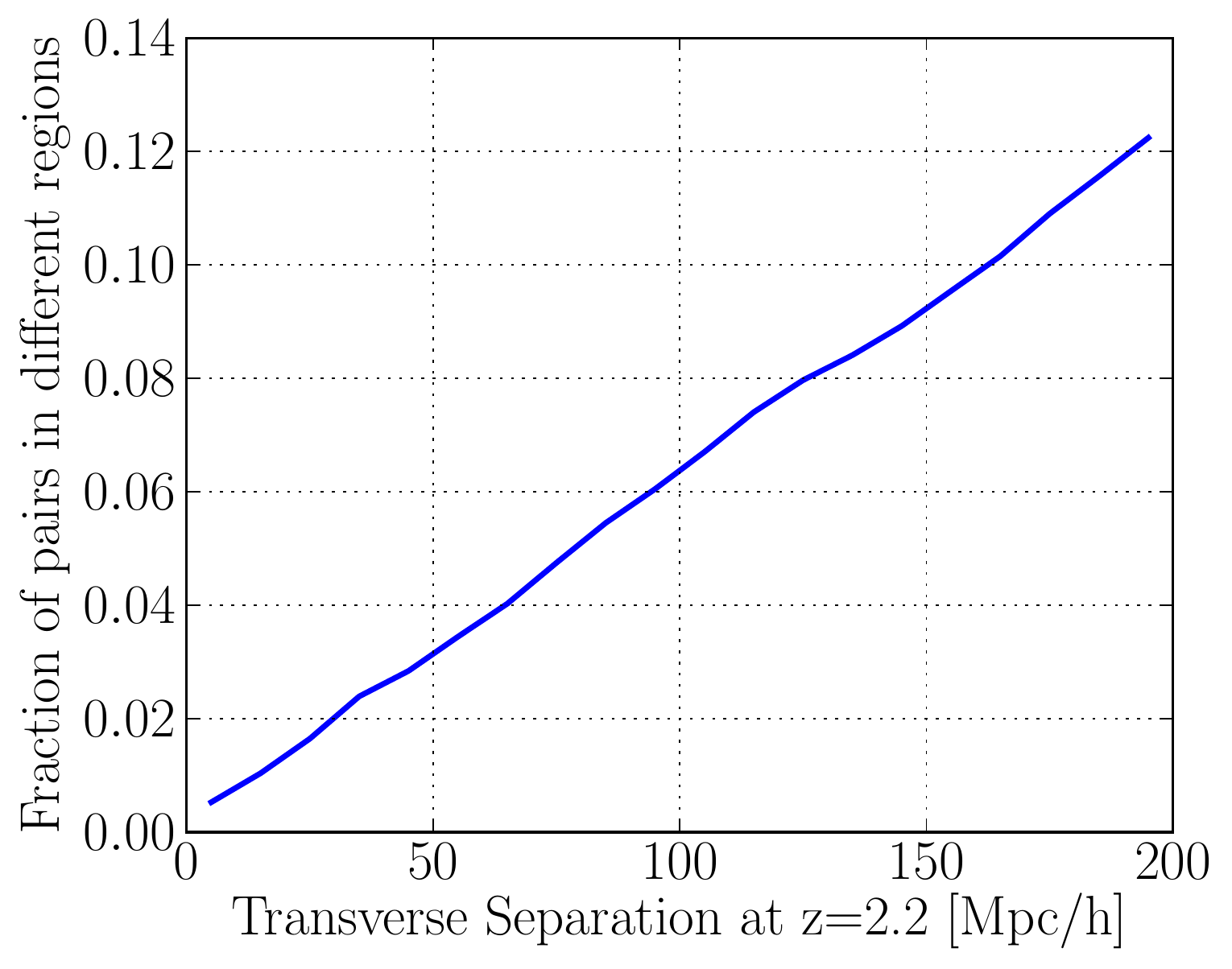}
 \end{center}
 \caption{Left panel : Independent regions of sky used to generate the DR11 
  \lya\ mocks.  
  Right panel: Fraction of QSO pairs that have their two members from different 
  regions as a function of transverse separation at redshift $z$=2.2.}
 \label{fig:chunks}
\end{figure}

As discussed above, the computational constraints require that the
data sample be split into regions with lower numbers of quasars.
The DR11 regions were selected by an iterative optimization code designed 
to make physically compact regions with approximately the same number of QSOs.
These regions are shown in the left panel of figure \ref{fig:chunks}.

Because the regions are treated independently, the procedure generates no 
correlations between neighboring quasars that happen to fall in different 
regions.
The right panel of Fig.~\ref{fig:chunks} shows the fraction of QSO pairs
which come from different regions as a function of the transverse
separation of \lyaf\ pixels at redshift $z=2.2$. 
For the 3D correlation function, this corresponds to an effective 
number of pixel pairs in the mocks that is lower than in the data, and this 
fact needs to be taken into account when comparing mock with data measurements.

\section{Generation of expanded mock spectra}
\label{sec:expanded}

In the previous section we described the process of generating the
``raw'' mock spectra consisting of the 
correlated transmitted flux fraction $F_q(\lambda)$ 
along quasar lines-of-sight as defined by eqn. \ref{eq:deltafield}.
In this section we describe the production of so-called 
``expanded mock spectra'' 
which superimpose the raw spectra on quasar continua and
then add astrophysical features such aa high column 
density absorbers and metal absorption lines, 
as well as instrumental features such as 
noise, spectro-photometric calibration errors and sky subtraction residuals.

%These mocks will be publicly released. Access and details on how to use them
%are described in Appendix~\ref{app:access}.

The mock data are intended to mimic BOSS \emph{coadded} spectra, which result
from the addition of several successive 20-minute individual exposures of the 
same object, in a given plate and a single 
night\footnote{Under normal observing conditions, the number of exposures 
for a field is increased until the signal-to-noise of the coadded image
reaches a target value.}. 
Most
of our discussion below refers to coadded spectra except for the section on
noise properties where a distinction between the noise in each exposure
and the noise in the coadds is necessary.

%%%%% Making lambda explicit
For a given quasar, the generated flux $f(\lambda)$ is 
(dropping the quasar index $q$ for clarity)  
\begin{equation}
f(\lambda) = \left\{ \left[ F(\lambda) \cdot C(\lambda) \right] \ast 
	\tilde{W}(\lambda,R_p,R_w) 
	+ N(\lambda) \right\} \cdot M(\lambda) 
	+ \delta f_\mathrm{sky}(\lambda) \;.
	\label{eq:finalmockflux}
\end{equation}
The parameters
on the right-hand side are as follows: 
$F(\lambda)$ is the  
transmission fraction as defined by eqn. \ref{eq:deltafield}
except that it is set to 1 outside the \lyaf;
$C(\lambda)$ is the PCA-generated quasar continuum;
%the $\ast$ symbol 
%denotes a convolution in wavelength space, 
$\tilde{W}(\lambda,R_p,R_W)$ is the Fourier transform of the BOSS 
resolution and pixelization kernel (eq.~\ref{eq:kernel}) 
which is convolved with the product $F(\lambda) C(\lambda)$;
%which depends on the wavelength dispersion $R_w$ 
%of the spectrograph and the pixel width $R_p$ 
%(both approximated as independent of $\lambda$), 
$N(\lambda)$ is the noise computed from our model (eq.~\ref{eq:noiseModel}); 
$M(\lambda)$ is a linear function of $\log(\lambda/1~\mathrm{nm})$ (eq.~\ref{eq:flux_miscalib})
used 
to ensure that each  mock spectrum has the same mean flux
and spectral index as the corresponding real spectrum; 
and $\delta f_\mathrm{sky}(\lambda)$ is the added sky subtraction 
residuals (fig.~\ref{fig:missky}). 

The procedure for generating the correlated \lya\
transmission field yields a mean absorption
and variance given by equations \ref{eq:meanabsorption}
and \ref{eq:absorptionvariance}.
That transmission field models 
regions of optically thin neutral hydrogen absorption, ignoring
high-density systems and non-\lya\ absorbers. 
In the following two subsections, we describe how we add to $F(\lambda)$ these
additional absorbers.
In the subsequant subsections we describe how we generate
the other factors in (\ref{eq:finalmockflux}) necessary to produce
the expanded mock spectra.

\subsection{High column density (HCD) systems}
\label{sec:DLAs}

Dense systems with high neutral hydrogen column
density produce wavelength intervals of complete
absorption surrounded by damped wings.  These structures affect the
measured \lya\ transmission correlations in two ways.
First, they affect the size of the \lyaf\ fluctuations directly impacting
the variance in the resulting long-range 3D correlations. 
Second, since these
systems are themselves biased differently than the optically thin regions,
they will also affect these correlations themselves.

Damped \lya\ systems (DLAs) have strong damped wings that allow for their easy 
identification, but Lyman-limit systems (LLS) of lower column density can 
also affect the correlations even if their damped wings are weak and 
individually not detectable.

We insert 
both LLS and DLAs with neutral hydrogen column densities~$N_\mathrm{HI} > 10^{17.2}\cm^{-2}$,
which we collectively designate as
high column density (HCD) systems, following the procedure that is 
described in \cite{2012JCAP...07..028F}. 
In brief, HCD systems are distributed only in pixels where the transmission $F$ 
is lower than a certain threshold $F_0$, defined such that the probability 
to have an optical depth $\tau$ larger than $\tau_0 = -\ln(F_0)$ is 1\%. 
The column density of the HCD systems are randomly drawn from an analytical model 
\cite{2002ApJ...568L..71Z} calibrated to match observations 
\cite{2009A&A...505.1087N} from SDSSII-DR7. 
Voigt profiles \cite{olivero_empirical_1977} are included in these regions assuming a constant Doppler 
parameter $b_D = 70~\kms$.

The effect of HCD systems on the \lyaf\ correlation function and the measurement 
errors was studied previously in \cite{2012JCAP...07..028F}. 
In section \ref{sec:tests} this effect is measured in our mock 
catalogs taking into account BOSS spectroscopic characteristics.

%--- Metals ---
\subsection{Metals}
\label{sec:exp_Metals}

In addition to absorption from hydrogen, metals in the intergalactic
medium can also absorb quasar light at discrete wavelengths inside the \lya\
forest.  Metal absorption lines are individually indistinguishable from \lya\ forest absorption
(at BOSS resolution and noise levels). 
These metals  add 
``unwanted''
correlations to our \lya~forest data in two ways. 
First, absorption correlations are
imprinted in individual  spectra by gas at a given reshift absorbing
at more than one wavelength. 
For example,
a gas system with  \lya\ absorption at wavelength $\lambda$ will have correlated
metal absortption at  wavelength
$\lambda \lambda_\mathrm{Met} /121.6 \mathrm{nm}$,
where $\lambda_\mathrm{Met}$ is the wavelength of any other transition.
While this type of correlation is most important within individual spectra,
3D correlations are also generated because neighboring lines of sight
sample redshift-correlated gas structures. 
%The presence of
%accompanying absorption at fixed wavelength 
%ratios adds line-of-sight correlations,
%that may lead to spurious 3D correlations. 
The second type of correlation results from the fact that all 
metal species are themselves tracers of large-scale structure
\citep{2014MNRAS.445L.104P}.
This leads directly to a
weak 3D metal autocorrelation superimposed on the dominant \lya~autocorrelation.
We focus here on the former effect and will 
explore the latter in a future publication.

We added metal absorption with a procedure that assumes that
all significant metal absorption is associated with strong
\lya~absorption.
For each mock spectrum, we considered pixels for which
the \lya\ transmission
$F(\lambda)$ was below 0.4.
For these cases, we decrease the flux by $\delta F_\mathrm{Met}$
of their corresponding flux bin at wavelength
$\lambda \lambda_\mathrm{Met}/121.6\mathrm{nm}$ for each metal line.

%Application of this method requires a list of significant metal lines
%and appropriate values for the absorption, $\delta F_\mathrm{Met}$.
%This was done following the procedure of  \cite{2011JCAP...09..001S}
%where quasar spectra were stacked after shifting each spectra so that it 
%is centered on a wavelength of high absorption.
%In addition to the
%strong absorption at the center, 
%the stacked spectra have
%absorption lines  at wavelengths
%separated  from the center by $\lambda_i-\lambda_0$ where
%$\lambda_0$ is the wavelength of absorption by the species responsible
%for the strong absorption and $\lambda_i$ is the wavelength of
%absorption of any other absorber.
%With these stacks,
%we identified 21 correlated absorption features, 9 of them corresponding 
%to \lya\ -metal correlations and the others to metal-metal correlations 
%(arising from the strong absorber being due to a  
%metal line rather than to \lya\ ).  
%We designate these metal-metal correlations as ``shadows'' hereafter 
%\cite{2010ApJ...724L..69P}. 
%These measurements are available as part of the MockExpander package 
%(see Appendix A).

Application of this method requires a list of significant metal transitions
and appropriate values for the absorption flux decrement, $\delta F_\mathrm{Met}$.
This was done following the procedure of  \cite{2011JCAP...09..001S}
where quasar spectra were stacked after scaling the
wavelength of each spectrum
so that selected strong absorption is centered at the restframe \lya\ transition.
The stacked spectra then have, necessarily, strong absorption at the
$121.6 \mathrm{nm}$, but also
absorption lines  at wavelengths
separated from this feature by $(\lambda_i /\lambda_0 )121.6 \mathrm{nm}$, where
$\lambda_0$ is the wavelength of absorption by the species responsible
for the strong absorption and $\lambda_i$ is the wavelength of
absorption of any other absorber.
With these stacks,
we identified 21 correlated absorption features, 9 of them corresponding
to \lya\ -metal correlations and the others to metal-metal correlations
(arising from the strong absorber being due to a 
metal line rather than to \lya). 
We designate these metal-metal correlations as ``shadows'' hereafter
\cite{2010ApJ...724L..69P}.
We allowed these flux decrements to vary by performing this procedure for
7 selected bins in simulated \lya\ forest flux.
%\textbf{
Measurements of these metal transition flux decrements%}
are available as part of the MockExpander package
(see Appendix A).

Using the line catalog and the deduced flux decrements, we added
metal lines (including shadow lines) to the mock spectra.
We did this by reducing the flux at 21 separate wavelengths for every strongly
absorbed \lya\ forest pixel. The introduction of shadow lines in
this way is unphysical but  this method is employed in order to reproduce the full set of 1D absorption correlations in the mock spectra as tested by stacking both mock and observed spectra (see section \ref{sec:test_metals}). Our algorithm for the addition of metal absorption assumes that all metal absorption can be characterized by its association with \lya\ forest absorption. The need for these shadow lines indicates that this is not always the case; some strong metal lines occur where little or no Lyman alpha absorption is apparent.

The ascribed metal absorption is fully determined by the mean metal absorption
above and so no scatter is added to these flux decrements. This will be refined
in a future publication.

\subsection{Quasar continua $C_\lambda$}
\label{sec:continua}

\begin{figure}[t]
 \begin{center}
  \includegraphics[width=0.6\textwidth]{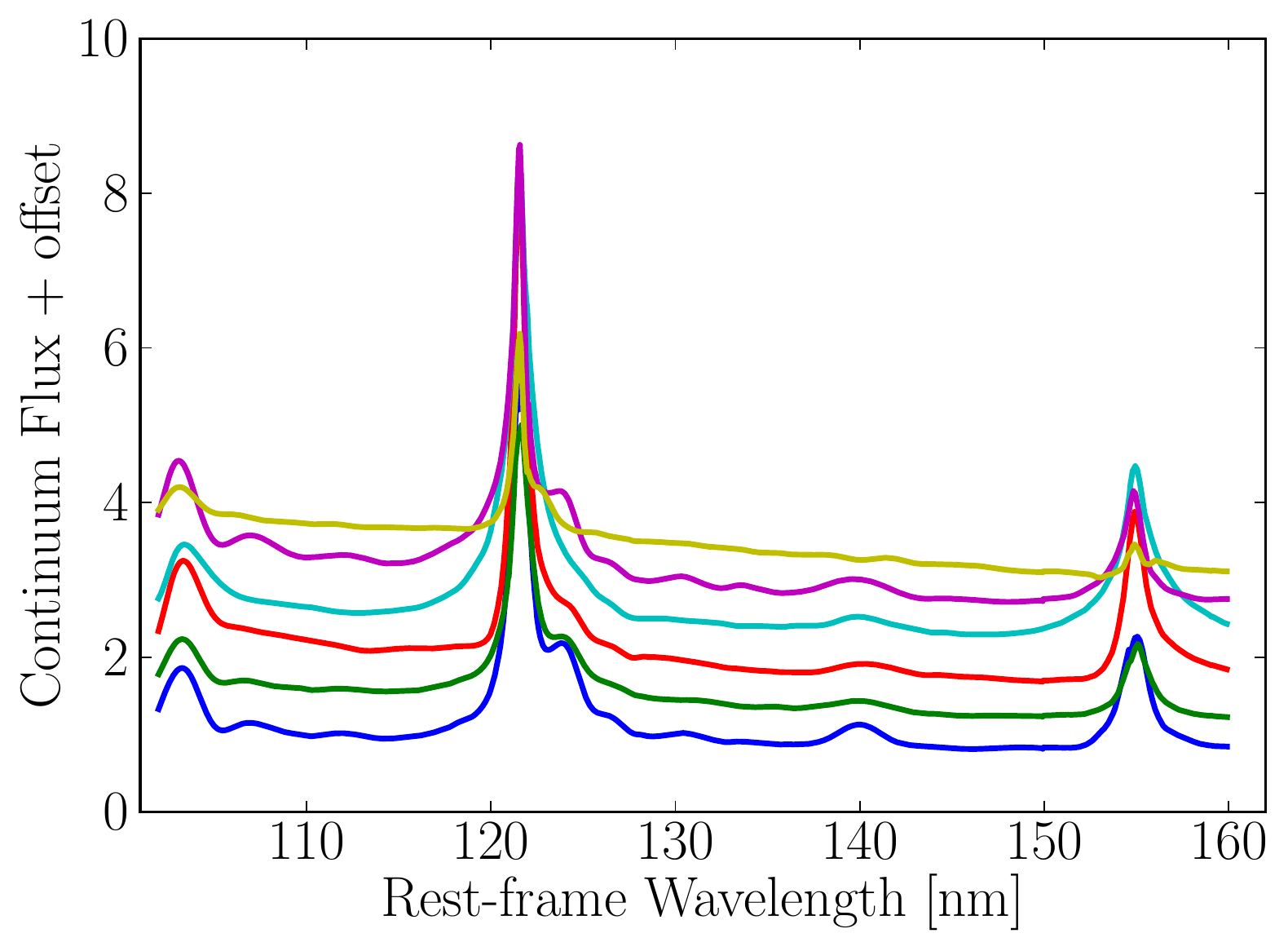}
 \end{center}
 \caption{Example continua from 5 random mocks showing variations in line 
  shapes and the continuum shape within the \lyaf.}
 \label{fig:continua}
\end{figure}

The quasar continuum $C_\lambda$ for each line-of-sight is constructed using eight 
higher ranked eigenspectra of the Principal Component Analysis (PCA) of  
low redshift quasar spectra\footnote{Low redshift quasars have less absorption in the forest, simplifying the continuum estimation.} \cite{2005ApJ...618..592S}. 
The final continuum is a sum of a mean shape and a linear combination of these 
eigenspectra for which amplitudes were randomly sampled following a centered 
Gaussian distribution with the corresponding  standard deviation.
These eigenspectra cover restframe wavelengths 102--160~nm spanning
from the Lyman-$\beta$ peak through C IV $\lambda 1549$.
Wavelengths above and below this range are discarded in the mock spectra.
Figure~\ref{fig:continua} shows 5 examples of mock continua with variations
in line shapes and the shape of the continuum within the \lyaf.
The random sampling of the PCA eigenvalue amplitudes occasionally leads to 
negative continua at some wavelength bins (this happens on 0.5\% of the 
continua), and in this case the continuum is discarded and a new set of random 
amplitudes is drawn.

\subsection{The BOSS kernel}

BOSS spectrographs cover the wavelength range 361 nm - 1014 nm with a 
resolving power $\lambda/\Delta \lambda$ varying from $1300$ in the blue end 
to $2600$ in the red end. 
Each data spectrum has its own estimate of the wavelength dispersion per pixel.
In BOSS coadded data, the pixels are logarithmic in wavelength with steps 
of $\Delta \log_{10}(\lambda/1~\mathrm{nm}) = 10^{-4}$ corresponding to $69~{\rm km\,s^{-1}}$.

As described in section~\ref{sec:absorption}, the raw absorption fields were produced 
over the lines-of-sight using a grid in comoving space with bins of 0.5$\hMpc$. 
To match BOSS spectra binning and resolution, we first compute the mean 
wavelength dispersion (PSF) $R_w$ and mean pixel width $R_p$ over the \lyaf\ region of 
the corresponding data spectrum. 
Each raw field was then convolved using the following kernel 
(in Fourier space):

\begin{equation}
W(k,R_p,R_w) =  \exp{\left( - \frac{k^2 R_w^2}{2}\right)} 
				\left[\frac{ \sin{(kR_p/2)} }{kR_p/2} \right].
\label{eq:kernel}
\end{equation}

We then match the binning by taking, for each data pixel, the absorption 
value of the closest pixel of the smoothed raw field.

\subsection{Flux normalization $M_\lambda$}
\label{sec:flux_miscalib}

We wish to ensure that each mock quasar has a mean forest flux and spectral
index equal to those of the corresponding real quasar.
Specifically, we normalize the noise-free mock quasar flux $f_{\rm mock}$ to 
the data flux $f_{\rm data}$ by fitting for $M_0$ and $M_1$ in
\begin{equation}
    f_\mathrm{\rm data}(\lambda) = 
f_\mathrm{\rm mock}(\lambda) (M_0 + M_1 \log_{10}(\lambda/1~\mathrm{nm})),
    \label{eq:flux_miscalib}
\end{equation}
over the rest-frame wavelength ranges
$104.1 < \lambda < 118.5$~nm (inside the \lyaf) and
$127.0 < \lambda < 150.0$~nm (between the \lya\ and CIV emission peaks). 
Then, $f_{\rm mock}$ is multiplied by the factor 
$M_\lambda=M_0 + M_1 \log_{10}( \lambda/1~\mathrm{nm})$.
%to generate mocks that match 
%the distortion in the data. 
%This also implies that some power on very large scales in the mocks is being 
%removed by this operation, but something similar occurs when matching the 
%data to a continuum template. 
These fits are done using the inverse variance given by the pipeline 
as fit weights and ignoring all masked pixels. For DR9 mock data-sets, 
the fit was performed without weighting 
leading a slightly larger number of bad fits, about 1\% of the full sample. 

The parameter $M_1$ effectively corrects for quasar spectral distortions
introduced by the SDSS optics that are currently not corrected by 
the pipeline.
The Sloan 2.5-m telescope has a chromatic focal plane and lacks an 
atmospheric dispersion corrector.  
As a result, the optimal position for a spectrograph fiber is a wavelength 
and airmass dependent quantity.
Galaxy targets and calibration stars are optimized for 540~nm, while quasar 
targets are offset both along and across the focal plane to optimize the 
signal-to-noise at 400~nm\ for \lyaf\ studies.
This offset means that the flux calibration vectors derived from the standard 
stars are not correct for the quasars and result in a flux mis-calibration 
which depends upon wavelength, airmass, seeing, guiding, and the location 
on the focal plane \cite{2013AJ....145...10D}.

%In Fig.~\ref{fig:Miscalibs} some examples of flux mis-calibrations are 
%shown (blue solid lines).

%\begin{figure}[t]
%\centering
%\includegraphics[scale=0.5]{Miscalibs.pdf}
%\caption{
%Comparison between flux mis-calibration (solid blue lines) and the noise 
%mis-calibration (dashed red line) used on the generation of a random set 
%of 20 mock spectra. 
%The former is obtained by matching mock and data flux values using 
%Eq.~\ref{eq:flux_miscalib}. 
%The latter is the noise mis-calibration amplitude coming from the ratio of 
%wavelength bin sizes between individual and coadded spectrum.}
%\label{fig:Miscalibs}
%\end{figure}

%Cosmological analyses which use BOSS data should fit the \lyaf\ continuum in 
%a way that marginalizes out any flux mis-calibration.
%Including this mis-calibration in the mock data-sets enables analyses to test 
%their robustness with respect to this effect.

\subsection{The noise, $N_\lambda$}
\label{sec:noise_model}

The noise, $N_\lambda$, added to the fluxes of a given
mock quasar  is a random number taken
from a Gaussian distribution of mean zero and with a variance determined by the
noise model for the corresponding real quasar.
The noise models are most naturally expressed using
the total number of photo-electrons $p_\mathrm{tot}$ in a given pixel (object plus sky),
since in an ideal system the variance would be equal to $p_\mathrm{tot}$ (pure Poisson noise).
In practice, our model gives the variance $\sigma^2_\mathrm{phot}$
as a linear function of $p_\mathrm{tot}$:
\begin{equation}
    \sigma^2_\mathrm{phot} = N_0 + N_1 p_\mathrm{tot}
	\label{eq:noiseModel}
\end{equation}
The coefficient $N_0$ reflects the CCD readout 
noise and other systematic effects that are independent of the photon flux. 
The coefficient $N_1$ would be unity for pure Poisson photon noise in the 
absence of systematics, but in practice $N_1 \geq 1$ owing to sky subtraction 
and flux calibration errors. 

\begin{figure}[t]
\centering
\includegraphics[scale=0.7]{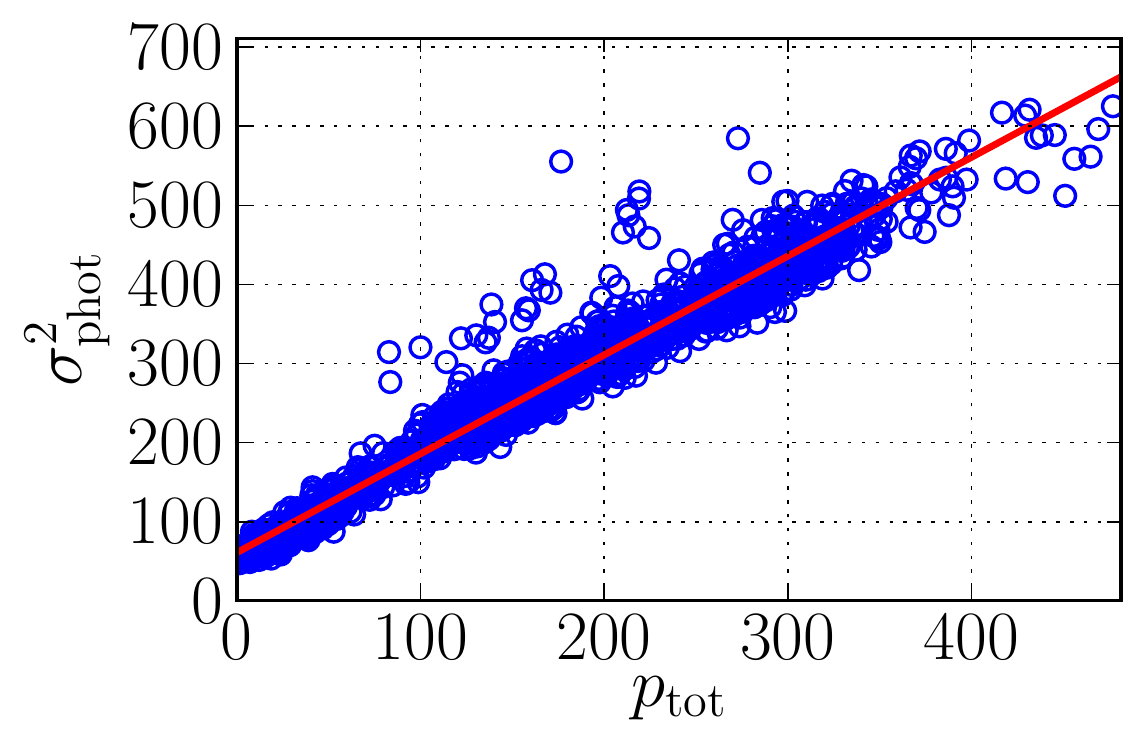}
\caption{
Example of a noise model fitting for spectrum 5493-56009-504 (plate-MJD-fiber).
The red line shows the result of the linear fit of the photon variance 
(Eq.~\ref{eq:noiseModel}) in the over data points ($N_0=68.7$ and $N_1=1.16$).
}
\label{fig:noiseModelFit}
\end{figure}

For each mock quasar, the parameters $N_0$ and $N_1$ are found by
fitting $p_\mathrm{tot}$ as a function of  $\sigma^2_\mathrm{phot}$
for the corresponding real quasar.  This requires the use of
a effective calibration vector $c(\lambda)$ to transform observed fluxes 
into photo-electrons:
\begin{eqnarray}
	\label{eq:fluxToPhot1}
    p_\mathrm{tot}(\lambda) & = & 
\left[f_\mathrm{QSO}(\lambda) + f_\mathrm{sky}(\lambda) \right] / c(\lambda) \\
    \sigma^2_\mathrm{phot}(\lambda) & = & \sigma^2(\lambda) / c(\lambda)^2
	\label{eq:fluxToPhot2}
\end{eqnarray}
where $\sigma^2(\lambda)$ is the estimated flux variance of the data.
We fit the linear noise model (Eq.~\ref{eq:noiseModel}) for each spectrum, 
using pixels from the blue side of the spectrograph.
Fig.~\ref{fig:noiseModelFit} shows an example of a fit for $N_0$ and $N_1$ for 
one quasar spectrum in DR11. 

In addition to $\sigma(\lambda)$, 
we also need the effective calibration vector $c(\lambda)$ in 
Eqs.~\ref{eq:fluxToPhot1}~and~\ref{eq:fluxToPhot2}. 
The BOSS data individual exposures include the calibration vector $c_i(\lambda)$
that converts between observed photo-electrons
$p_i(\lambda)$ and calibrated flux 
$f_i(\lambda)$. The coadd is performed as a weighted simultaneous spline 
fit to the individual exposures and the resulting effective calibration vector 
is not calculated.
We re-derive the effective calibration vector using the approximation that the 
coadded photo-electrons are the unweighted sum of the individual exposure photo-electrons:
$p = \sum_i p_i(\lambda) = r(\lambda) \sum_i f_i(\lambda)  / c_i(\lambda)$,
where $r(\lambda)$ corrects for the wavelength dependent difference in bin-size 
between individual exposures and the coadded spectra.

Since the coadded flux is normalized to the same units as the individual 
exposures (ergs/s/cm$^2$/\AA), we may factor out 
$f(\lambda) \simeq f_i(\lambda)$ such that
$p(\lambda) = f(\lambda) r(\lambda)\sum_i c_i^{-1}(\lambda)$.
Thus the effective calibration vector to convert between coadded photo-electrons and 
coadded flux is:
\begin{equation}
    c(\lambda) = \left( r(\lambda) \sum_i c_i^{-1}(\lambda) \right)^{-1}
    \label{eq:calibration_vec}
\end{equation}
For this calculation we use only the blue exposures, therefore limiting the 
wavelength range of mock spectra from $360$ to $633$~nm.
 
With the parameters $N_0$ and $N_1$ for each
quasar, it is simple to add realistic noise to the mock spectra. 
We compute the mock quasar flux from the product of the transmission $F$ and 
the generated PCA quasar continuum $C$, and add the same sky flux 
$f_\mathrm{sky}$ that is used in the data. 
We then compute the mock photons $p$ and noise $\sigma_p$ at each pixel,
\begin{eqnarray}
    p & = & (F \cdot C + f_\mathrm{sky}) / c \\
    \sigma^2_p & = & N_0 + N_1 p
\end{eqnarray}
We add noise using a Gaussian distribution with mean 0 and sigma $\sigma_p$:
\begin{equation}
    \tilde{p} = p + {\cal N}(0, \sigma_p)
\end{equation}
We convert back into quasar flux $\tilde{f}_\mathrm{QSO}$ :
\begin{equation}
    \tilde{f}_\mathrm{QSO}  = \tilde{p} c - f_\mathrm{sky}
    \label{eq:noisy_flux}
\end{equation}
\begin{equation}
    \sigma_{\tilde f}  =  c \sigma_p
\end{equation}

%The final products are the noiseless mock spectrum, a noisy realization of that 
%spectrum, the true inverse variance of that noise, 
%and a noisy measurement of that inverse variance to mimic the fact that the 
%real data measurement errors are themselves a noisy estimate of the true 
%errors.

The next step consists in mis-reporting the value of the 
true variances $\sigma^2(\lambda)$, 
as we observe in real data.
%Of course the accuracy of the noise model is only as good as
%the accuracy of $\sigma(\lambda)$.
%We do not use the pipeline-calculated $\sigma(\lambda)$ because there are
%systematic deviations between it and the true $\sigma(\lambda)$.
Biased estimates of pixel variances on real data were reported by different
studies \cite{2013A&A...559A..85P, 2013AJ....145...69L}. They find
%smooth spectral regions redwards of the \lya\ peak, between 142.0 and 151.0~nm
%where there are no strong emission lines. 
%For each QSO spectrum we fit a 3rd order polynomial in this region to
%obtain an estimate of the continuum.
%The polynomial fit includes 5-$\sigma$ outlier rejection to be robust against 
%metal absorption and data artifacts. We then organize the data in
%wavelength bins and measure, in each bin, the spread of the pull distribution 
%by fitting a Gaussian to this distribution in the range $[-2,2]$. 
%Deviations from unity of the sigma parameter of this Gaussian 
%directly measure any pipeline mis-estimation of the noise variance.
that for individual exposures, the pipeline noise 
estimates are accurate at the 1--2\% level.
Coadded spectra have a larger and wavelength dependent
error in the estimated noise.
For $\lambda <600$~nm (blue spectrographs), the biases are  
approximately proportional to the square root of the
ratio of the coadd to individual spectrum spectral bin sizes.
Below (above) $\sim$475.0~nm, where the coadded bins are smaller (larger) 
than the original 
exposures, the pipeline underestimates (overestimates) the noise by 0--10\% (0--15\%).
A flux-dependent bias of the pixel variances was also reported in \cite{2013AJ....145...69L}.
We use the observed wavelength dependent errors in the pixel variances to 
purposefully mis-report the noise in mock spectra. 

We also add Gaussian random fluctuations to this noise estimates with standard 
deviation proportional to the photon variance itself. The final mock photon 
noise estimate $\tilde \sigma_{\tilde f}$ is given by 
$\sigma_{\tilde f}(\lambda)r(\lambda) + {\cal N}(0, \sqrt{2}\sigma_f )$.

The model presented here assumes that the noise in different 
pixels of the same spectrum 
is uncorrelated. This is likely not realistic because covariance among 
neighboring pixels is introduced by rebinning. However, other sources of 
small scale correlations in these mock catalogs are not 
correctly modeled by the input power spectrum, since the log-normal model was 
built to fit large-scale correlations. 
Therefore we ignore any noise correlation between neighboring bins.  

An alternative noise model \cite{2014arXiv1405.1072L} uses four per-object 
parameters (instead of two in our model). This method solves simultaneously for both
the noise parameters and the optimal co-added flux. 
However, this modeling is inappropriate for our current purposes 
since our goal is to add noise to the mock co-added flux, which follows the pipeline and is not optimal.

%\af{Removed "we" before "our goal"}

%--- Sky Mis-subtraction ---
\subsection{Sky mis-subtraction}

\begin{figure}[t]
 \begin{center}
  \includegraphics[width=0.44\textwidth]{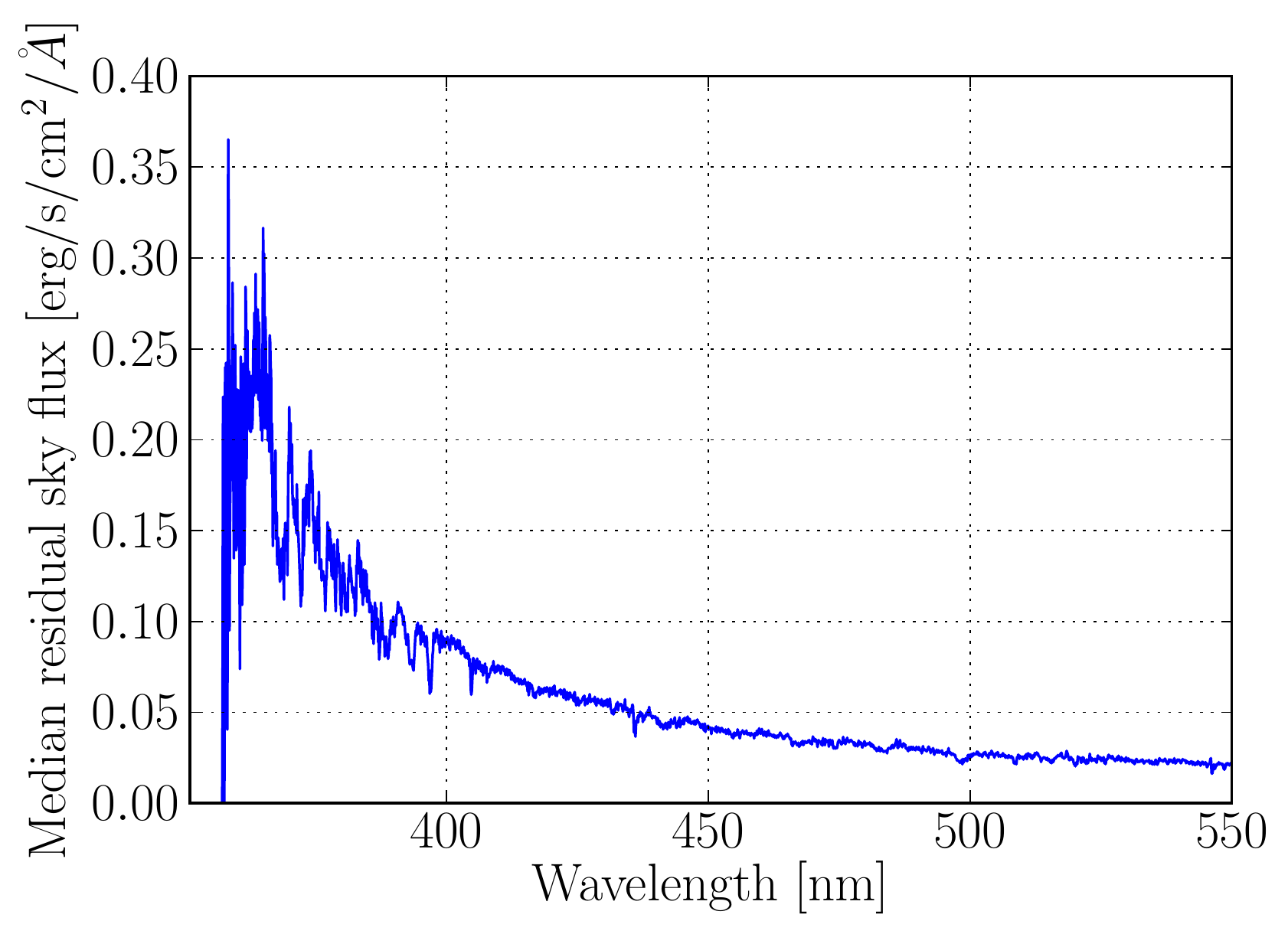}
  \includegraphics[width=0.44\textwidth]{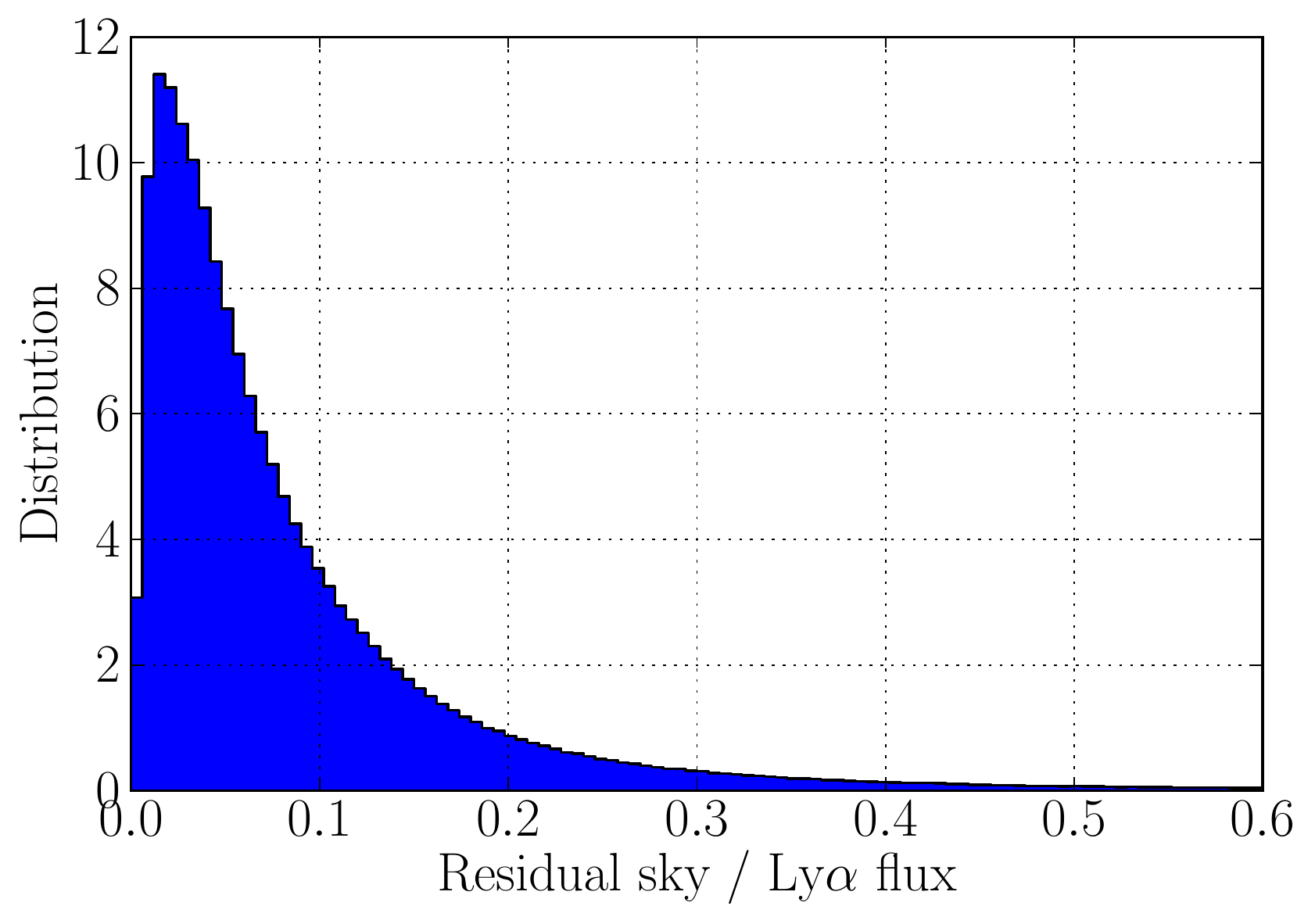}
 \end{center}
 \caption{Median residual from BOSS sky-subtracted sky fibers (left panel), 
 and the distribution of the ratio of that residual to simulated \lyaf\ 
 flux (right panel).}
 \label{fig:missky}
\end{figure}

Figure \ref{fig:missky}~(left) shows the median residual of BOSS sky spectra 
after the sky model has been subtracted.  While this is only a 1--2\% bias in 
the sky subtraction, it can be large compared to the \lyaf\ flux, which is 
typically faint compared to the sky.
Fig.~\ref{fig:missky}~(right) shows the relative flux between noise-free
simulated \lyaf\ flux and the residual sky.  
The median is $5.6\%$, the mean is $9.9\%$, and there is a tail reaching up 
to 1.0.
Note that this is an additive component to mock spectra, unlike other 
mis-calibrations that are multiplicative.

Mock spectra add this median sky subtraction residual times a random constant 
scatter with mean 1.0 and RMS 0.1 so that each spectrum receives a slightly 
different sky subtraction residual bias.

\section{Comparisons of mock and real spectra}
\label{sec:tests}

In this section we compare several properties of the mock spectra
to those of real spectra:
continuum characteristics, noise levels,
flux transmission correlations within individual spectra,
and absorption by metals. 
Unless otherwise stated, the standard set of mocks used for the comparisons in this section does not contain metals or high column density systems (HCDs), but includes all instrumental systematics. The effect of metals and HCDs is studied in the end of section~\ref{sec:3d_clustering}.

%\af{Within this section, when we say measurement on "mocks", are we including
%metals / HCD systems? It is not very clear...}

\subsection{Mean continuum, variance and diversity}
\label{sec:continuum_miscalib}

\begin{figure}[t]
 \begin{center}
  \includegraphics[width=0.6\textwidth]{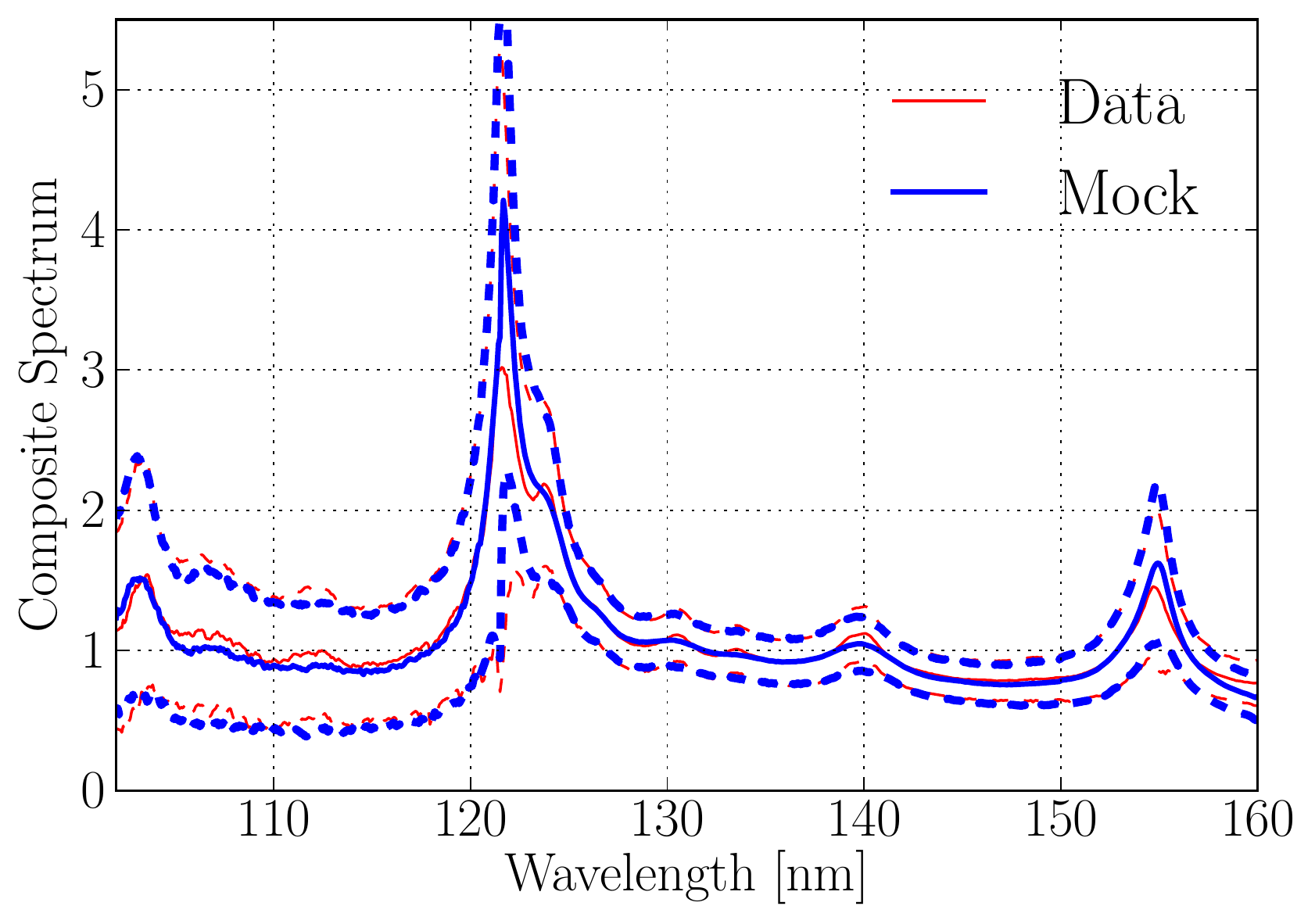}
 \end{center}
 \caption{Composite of 10,000 quasar spectra using data (thin red lines) and 
 mocks (thick blue lines). 
 The solid lines show the mean normalized flux values and dashed lines shows 
 the $\pm 1$-$\sigma$ region. 
 Stacked spectra are normalized to the mean flux observed in the rest frame 
 wavelength range 126 and 138~nm.}
 \label{fig:CompositeSpec}
\end{figure}

As a first test, we compare the mean spectra of the data and mocks.
This simple test verifies that the product of mean continuum and 
forest transmission is reasonable.
The mean spectra were calculated by first shifting each spectrum
to its rest frame and
normalizing it such that the average flux in the wavelength interval from 
126 nm to 138 nm is unity.
We stacked the set of normalized spectra to
obtain the mean product of continuum and transmission 
and its variance as a function of rest-frame wavelength.

The solid lines in figure \ref{fig:CompositeSpec} shows the resulting
mean spectra, red for the data and blue for the mocks. 
The dashed lines show the flux at one standard deviation higher
and lower than the mean.
The largest differences
are in the height of the \lya\ emission peak and 
to a lesser extent in the height
of the iron lines in the \lyaf.
Whether these differences are an issue or not depends on the
application. 
For instance, the methods in \cite{2011JCAP...09..001S,2013A&A...552A..96B, 2013JCAP...04..026S, delubac_baryon_2015} only use the \lyaf\ for 
the purposes of measuring the 3D correlation function, and the composite
spectrum is computed self-consistently (which would automatically account for 
the shape differences when analyzing mocks or data).

%\af{The DR9 publications also measure the composite spectrum self-consistently,
%right?}

\begin{figure}[t]
 \centering
  \includegraphics[width=0.49\textwidth]{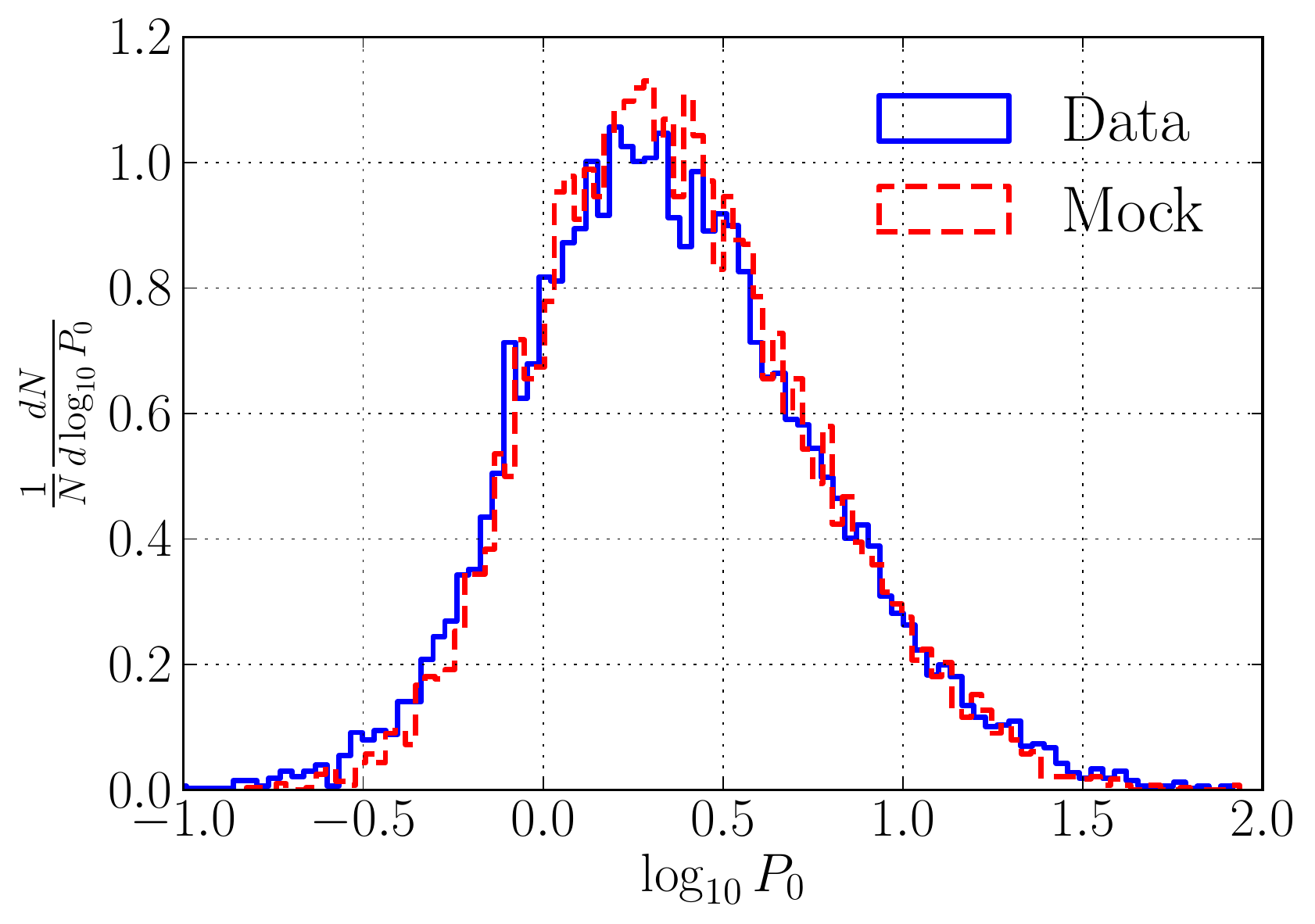}
  \includegraphics[width=0.49\textwidth]{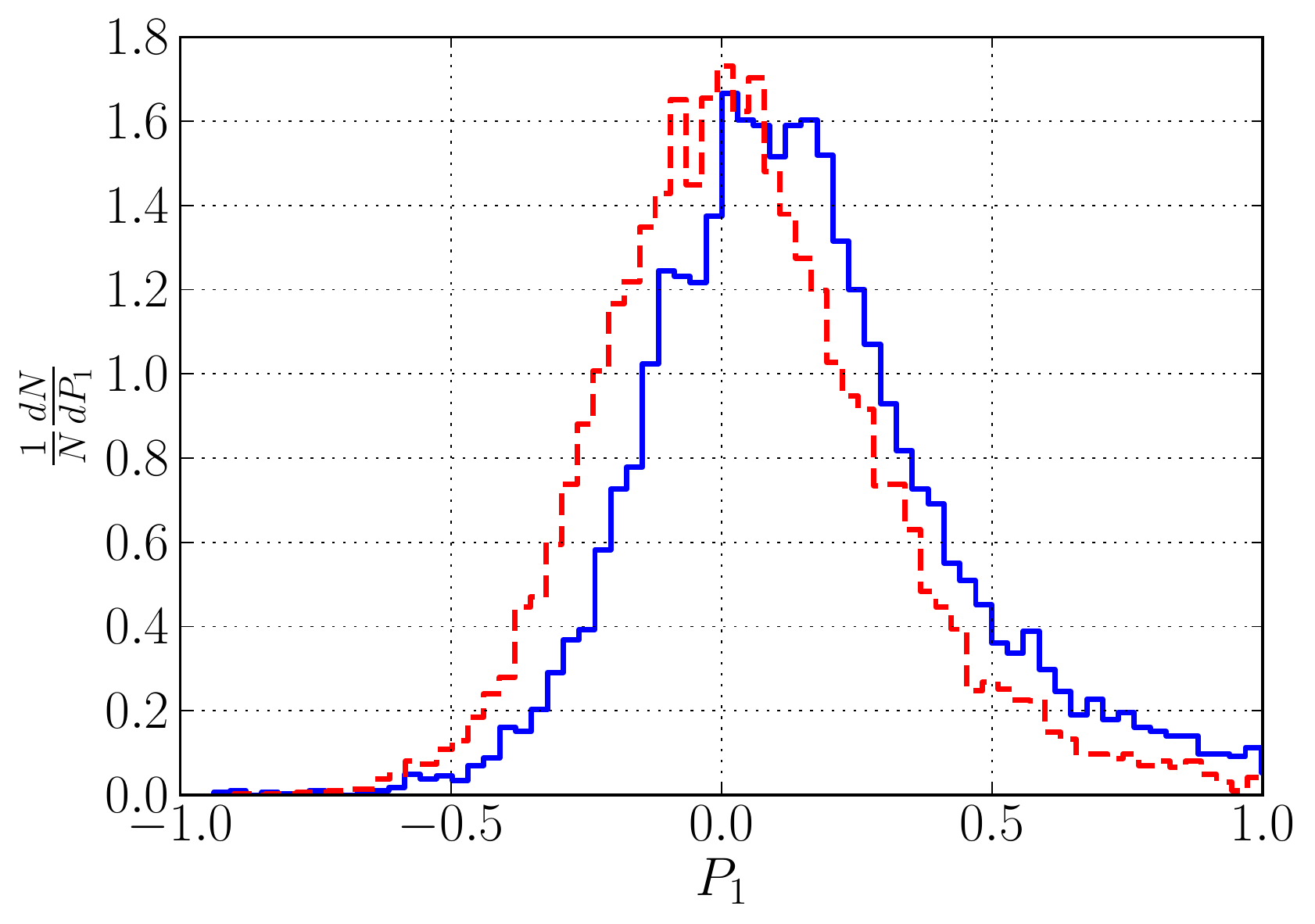}
 \caption{Distribution of amplitudes $P_0$ (left panel) and tilt 
 $P_1$ (right panel) in mocks (solid blue) and data (dashed red). 
 }
 \label{fig:contpars}
\end{figure}

We have also performed a second test of spectral diversity in the region of
the forest. For this test, we applied the continuum fitting C2 of 
\cite{delubac_baryon_2015} and 
%compare the distributions of the values of the
%best continuum fit parameters. Following \cite{delubac_baryon_2015}
and then fit the observed flux between the upper and lower limits
of the forest, $\lambda_0$ and $\lambda_1$:
\begin{equation}
f(\lambda)=\left(C_0{\lambda_1-\lambda\over\lambda_1-\lambda_0}
+C_1{\lambda-\lambda_0\over\lambda_1-\lambda0}\right)\bar{f}(\lambda_\textrm{rest})
\label{eq:cont_fit_model}
\end{equation}
where $f(\lambda)$ is the flux, $\bar{f}(\lambda)$ is the quasar-averaged flux,
$C_0$ and $C_1$ are free positive parameters (to ensure a
positive continuum).
The upper limit is
$\lambda_1=120(1+z_\textrm{qso})$ nm and $\lambda_0$ is the larger of 360 nm and
$104(1+z_\textrm{qso})$ nm. 
%The parameters $C_0$ and $C_1$ are the values
%of the fitted continuum at the ``edges'' of the forest for a given quasar.

For each forest we define a normalization parameter $P_0=(C_0+C_1)/2$ and
a tilt parameter $P_1=(C_1-C_0)/C_0$. The distributions of these two 
parameters are given in figure~\ref{fig:contpars}. There is an excellent
agreement in the distributions of $P_0$ and $P_1$ in mocks and data.
\textbf{The average value of the tilt
is slightly smaller on mocks due to a combination of factors affecting 
the continuum fitting estimates: differences in the sky residuals,
mean transmission and underlying 
transmission PDF. These factors are hard to estimate accurately from the data 
and to model in the mocks.} 
%However, the right panel shows that, on average, mocks
%have a slightly positive tilt while data have a slightly negative tilt.
%This fact might be due to our PCA components having been derived
%at much lower redshifts than the typical redshifts of our quasar sample.
%At any rate, this is 
%not a problem as long as the continuum fit model is flexible enough
%to compensate for that (as is the model in equation \ref{eq:cont_fit_model}). 
The distribution of amplitudes and tilts have similar
spreads in mocks and data, demonstrating that our mocks capture the
diversity of real data in the region of the \lyaf, at least regarding
the spectral index variability.

\begin{figure}[t]
\begin{center}
\includegraphics[width=0.5\textwidth]{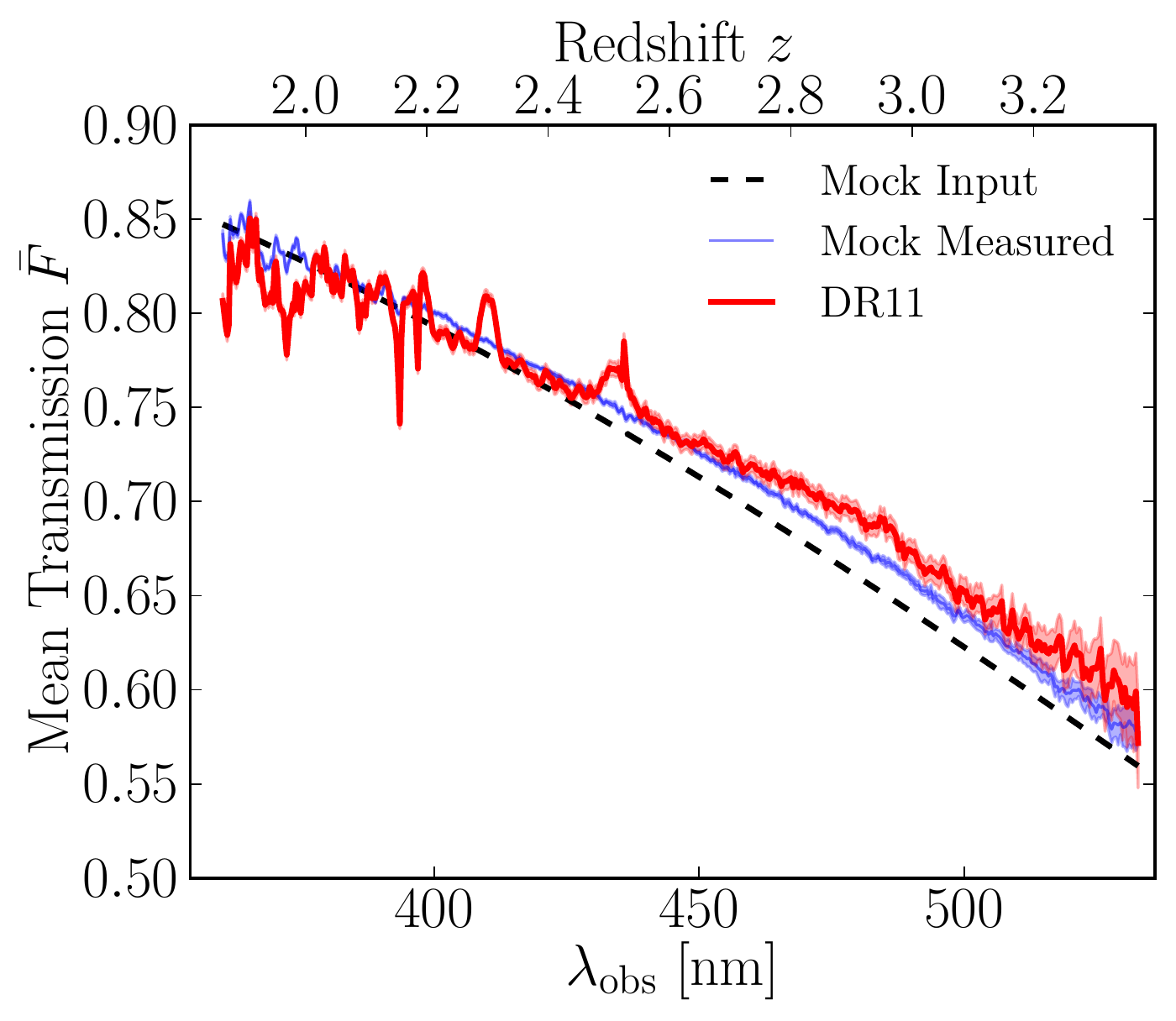}
\end{center}
\caption{
Mean transmission as a function of redshift
in mocks (blue) and data (red).
The black dashed line shows the $\bar{F}(z)$ from equation \ref{eq:meanabsorption}
that was used in the production of the mock spectra.}
\label{fig:mean_trans}
\end{figure}

The continuum fitting method C2 of \cite{delubac_baryon_2015} provides an
estimate of the unabsorbed continuum, $C_\lambda$.
Stacking spectra in observed wavelength then gives the mean transmission
as a function of redshift or, equivalently, of observed wavelength.
Figure \ref{fig:mean_trans} shows this quantity for the data and mocks.
Data reduction features are visible in the data stack (dashed red), such as 
the galactic calcium absorption (393.4 and 396.8~nm) or the Balmer 
residuals (near
398 nm, 410 nm and 435 nm). Some of those features are also present in the
stack of mocks (solid blue) as a result of the addition of the sky residuals
from figure \ref{fig:missky} but are not otherwise explicitly included.

\subsection{Noise}
\label{ssec:noise_props}

\begin{figure}[t]
\centering 
  \includegraphics[width=0.6\textwidth]{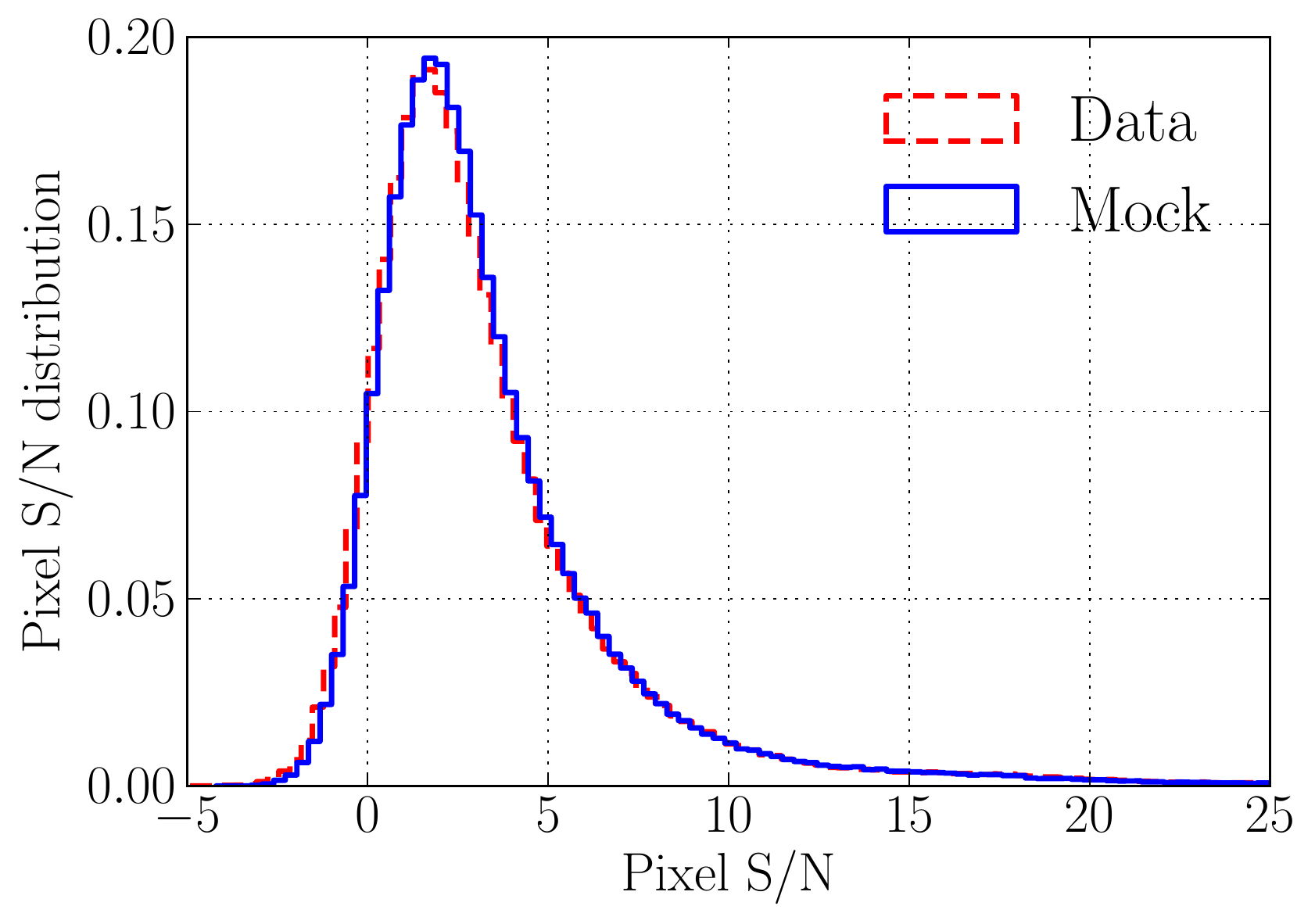}
 \caption{Distribution of \lyaf\ pixel signal-to-noise ratio in mocks 
 (solid blue) and data (dashed red).}
 \label{fig:SN_mocks_data}
\end{figure}

\begin{figure}[t]
 \begin{center}
   \includegraphics[width=0.45\textwidth]{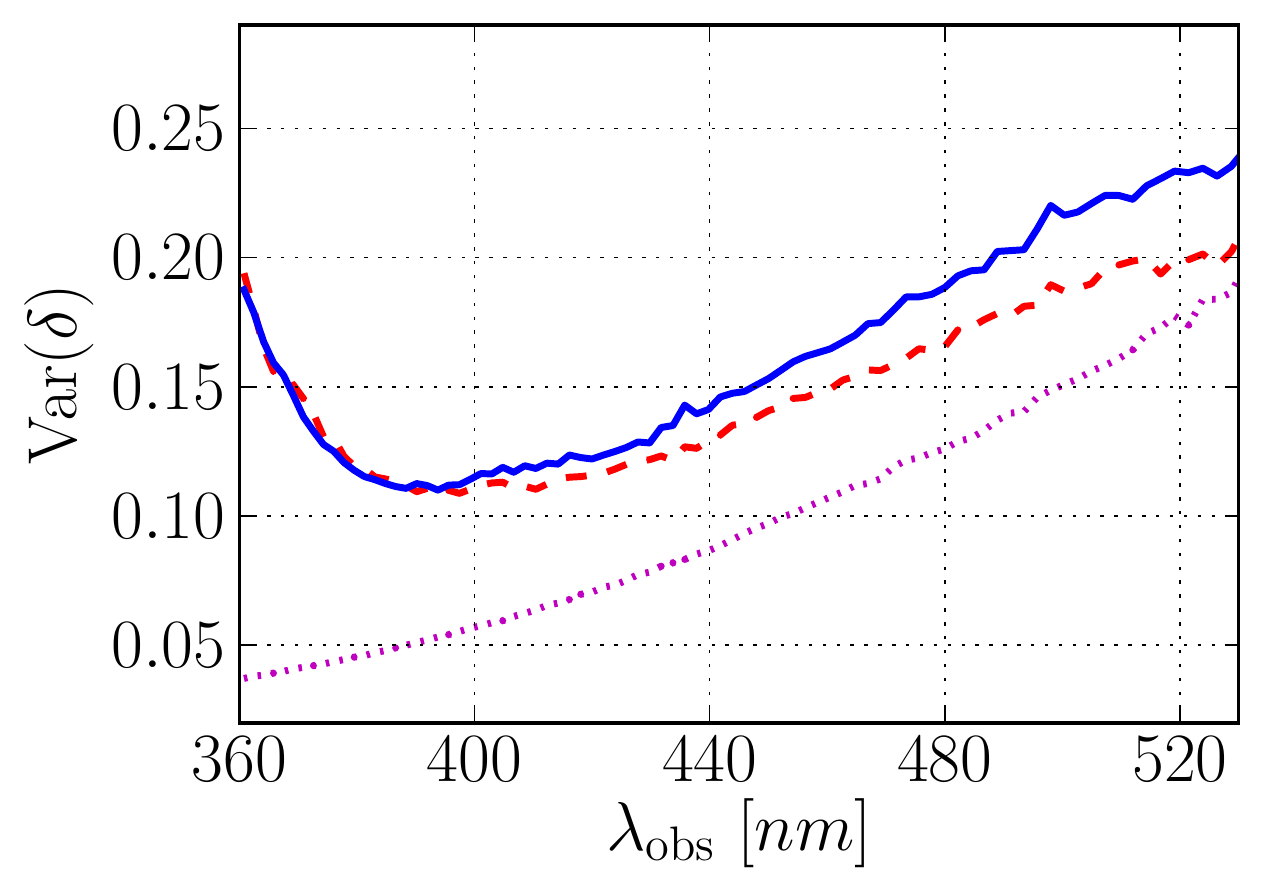}
 \end{center}
 \caption{ Variance of the flux transmission as a function of redshift 
for data (red dashed lines), 
expanded (blue solid line) and raw mock spectra (magenta dotted line). 
}
 \label{fig:mock_data_var1d}
\end{figure}

%In this section we compare the noise properties of mocks and data. 
The importance of accurately simulating noise is due to the fact that
noise accounts for about one half of the variance of individual
measurements of the flux transmission, the other half coming from
the intrinsic variations due to large-scale structure.
In turn, the variance of individual measurements accounts for
about one half of the variance of the measured three-dimensional
correlation function with the other half due to correlations
between measurements of neighboring pixels on individual spectra.
%This 
%comparison is important to validate our noise model and ultimately 
%demonstrate that the instrument and data reduction is sufficiently well
%understood.
%

%How closely the level of noise in mocks needs to reproduce that of real
%data depends on the application.
%In the case of the measurement of the 3D correlation function,
%since we approximate the noise as uncorrelated across \lyaf\ pixels
%(we discuss this approximation further on), the level of noise
%exclusively affects the variance of the measurement.
%For the noise level in BOSS data, the variance of the 
%3D correlation function has comparable contributions from large scale
%structure (LSS) and instrumental noise \citep{delubac_baryon_2015}, and the 
%contribution from both combined is about half of the total variance 
%(the rest coming from LSS correlations across \lyaf\ pixels in the
%same QSO).

%\af{That's a strange way to put it. I guess by LSS above you meant the 
%intrinsic variance in a pixel, right? When I read that, I thought you were
%talking about cosmic variance. The intrinsic variance in each pixel is just 
%the $r \rightarrow 0$ limit of the line of sight correlation, so making the 
%distinction is a little bit confusing... but probably adding a clarification
%is enough.} 

Figure \ref{fig:SN_mocks_data} compares the distribution 
of signal-to-noise ratio
for individual pixels in the forest in mocks and data. The signal-to-noise is
defined, for each individual pixel, as the ratio between the flux and the 
square-root of the pipeline estimate of the variance (before applying the 
corrections discussed in section \ref{sec:noise_model}). 
Given that in most 
studies the spectra are normalized to an unabsorbed continuum estimate, this 
ratio is the relevant quantity to compare. We find a very good agreement 
between mocks and data.

%Noise is a major component of the 
%variance of the $\delta$-field defined by equation
%(\ref{eq:deltafield}).
Figure \ref{fig:mock_data_var1d} 
shows the variance of the 
individual measurement of the transmission, $\delta_i$,
as a function
of redshift. It has contributions
from noise and large-scale-structure fluctuations. The latter contribution can
be computed using noiseless mocks, as is shown by the dotted magenta line.
It
represents roughly 50\% of the total variance.
The agreement between mock and data variances
is good (few percent level) below $\lambda=440$~nm, which corresponds to a redshift of 2.6, including a considerable fraction of analysis pixels 
\citep{delubac_baryon_2015}.
It is degraded by up to 10-15\% up above 480~nm ($z \sim 2.95$), beyond which there
are very few \lyaf\ pixels in the BOSS sample.

%\af{How did you choose 389 and 482? They seem pretty arbitrary. Why do you 
%highlight the mode, instead of mean/median? May be it would be nice to 
%mention also $z$, since this is what matters at the end?}

\subsection{Correlations within individual forests}
\label{ssec:xi1d}

\begin{figure}[bt]
 \begin{center}
  \includegraphics[width=0.45\textwidth]{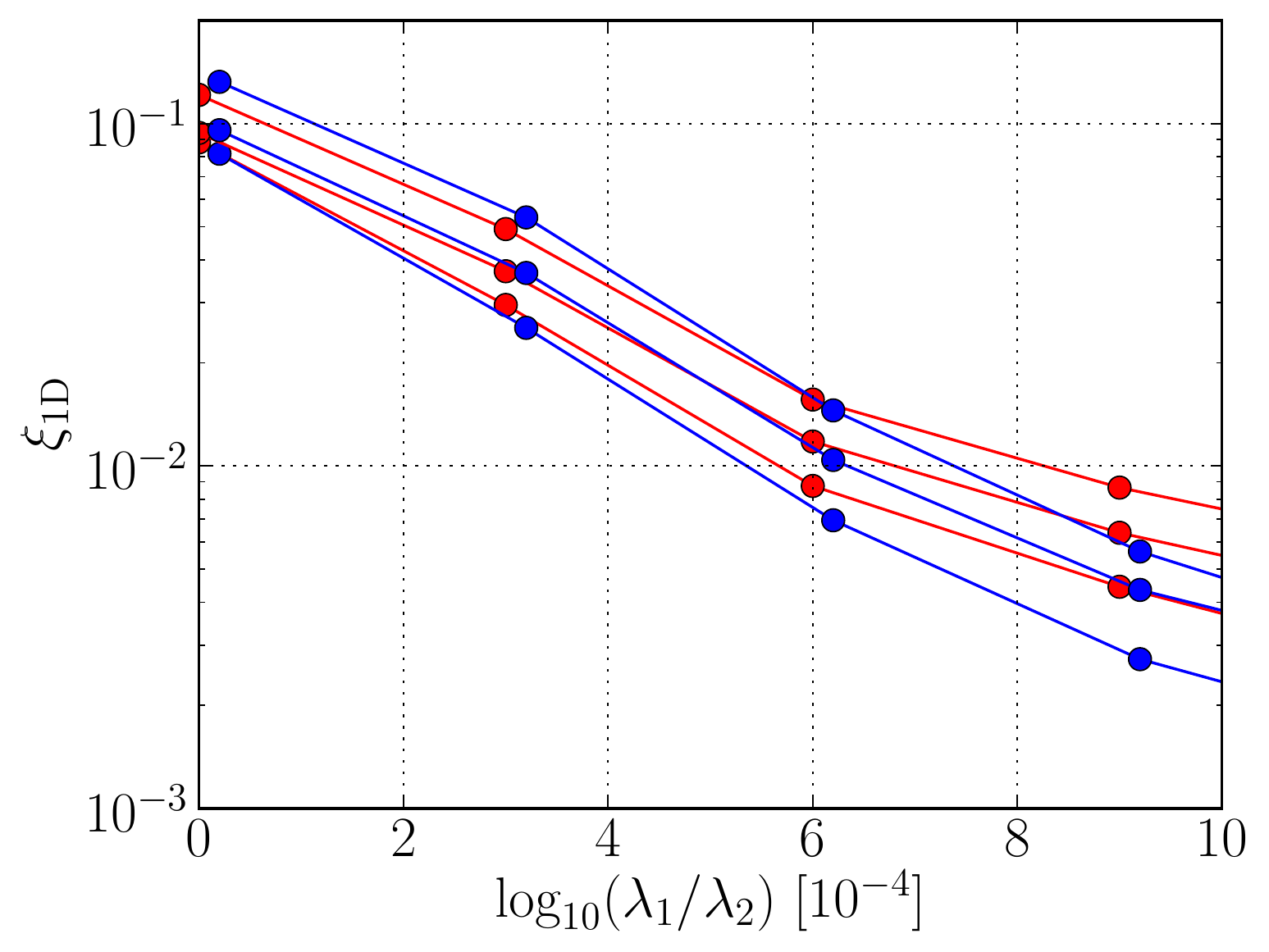}
  \includegraphics[width=0.45\textwidth]{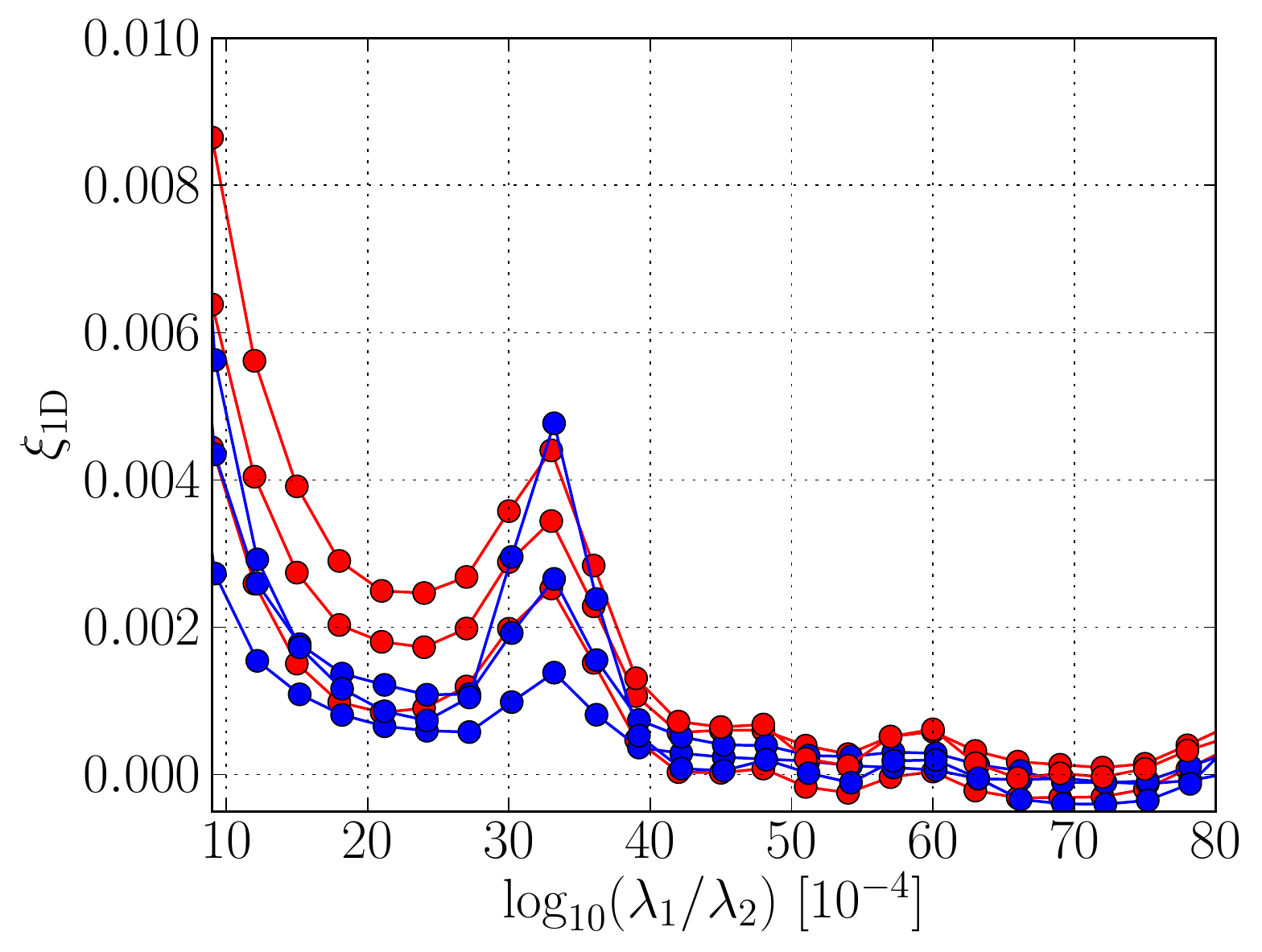}
 \end{center}
 \caption{ Correlation
$\xioned$ as a function of $\log_{10}(\lambda_1/\lambda_2)$ 
for mocks (blue points)  
and data (red points) 
for three bins in redshift:  
$2.0<z<2.3$; $2.3<z<2.6$; $2.6<z<2.9$ 
(by order of increasing correlations).
}
 \label{fig:mock_data_xi1d}
\end{figure}

Here we compare the correlations of $\delta$ within individual forests,
described by the correlation function:
\begin{equation}
\xioned(\lambda,\Delta\lambda) = \langle\delta_q(\lambda)\delta_q(\lambda+\Delta\lambda)\rangle
\end{equation}
for which we have used the same estimator (Eq.~\ref{eq:2ptcorrelation})
and weighting scheme as in 
\cite{delubac_baryon_2015} but changed the binning to work in wavelength instead
of comoving separation. For this comparison, we used mocks including metal absorption. 
As shown in figure \ref{fig:mock_data_xi1d},
$\xioned$ is a rapidly decreasing function of wavelength
separation except near $\log_{10}(\lambda_1/\lambda_2)\sim30\times10^{-4}$ where
it has a peak due to SiIII absorption correlated with \lya~absorption.

\begin{figure}[bt]
 \begin{center}
  \includegraphics[width=0.45\textwidth]{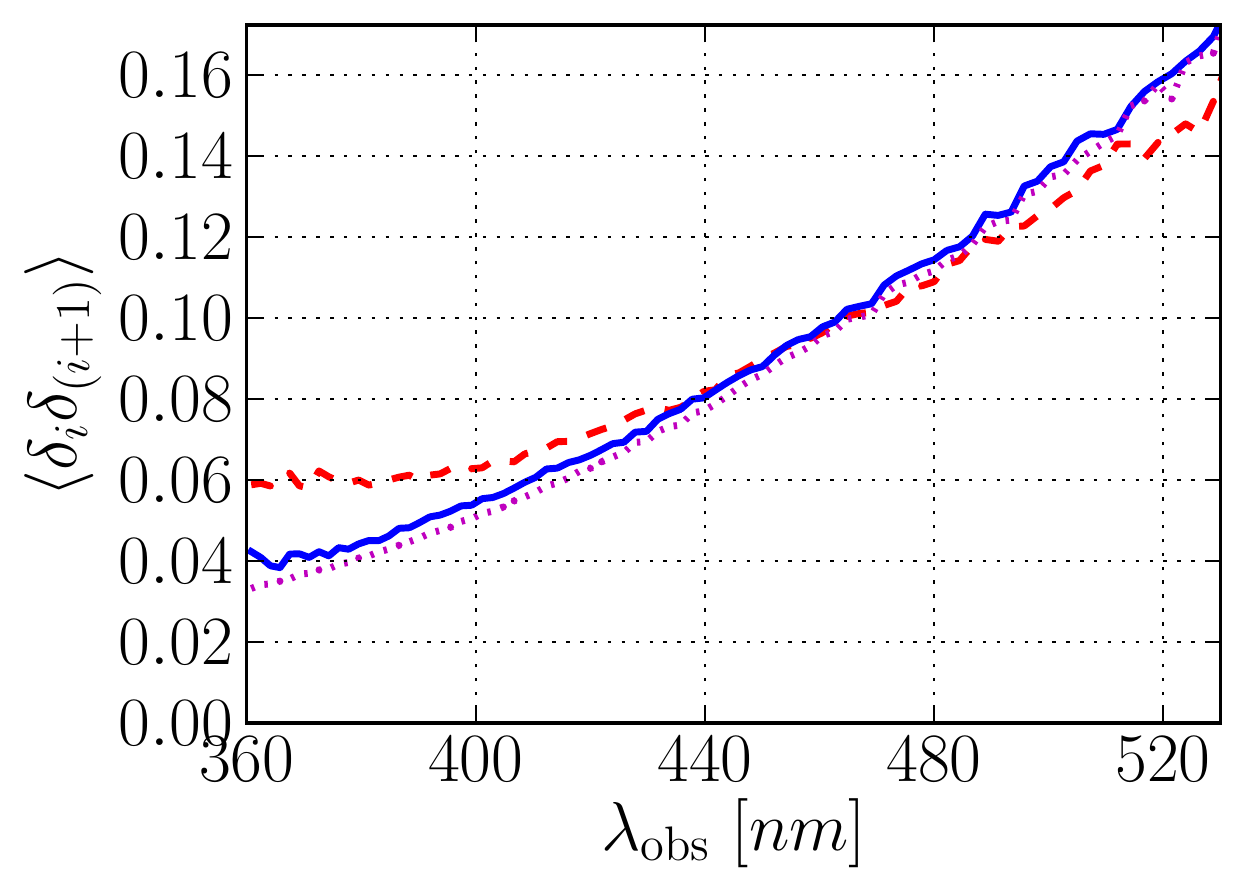}
  \includegraphics[width=0.45\textwidth]{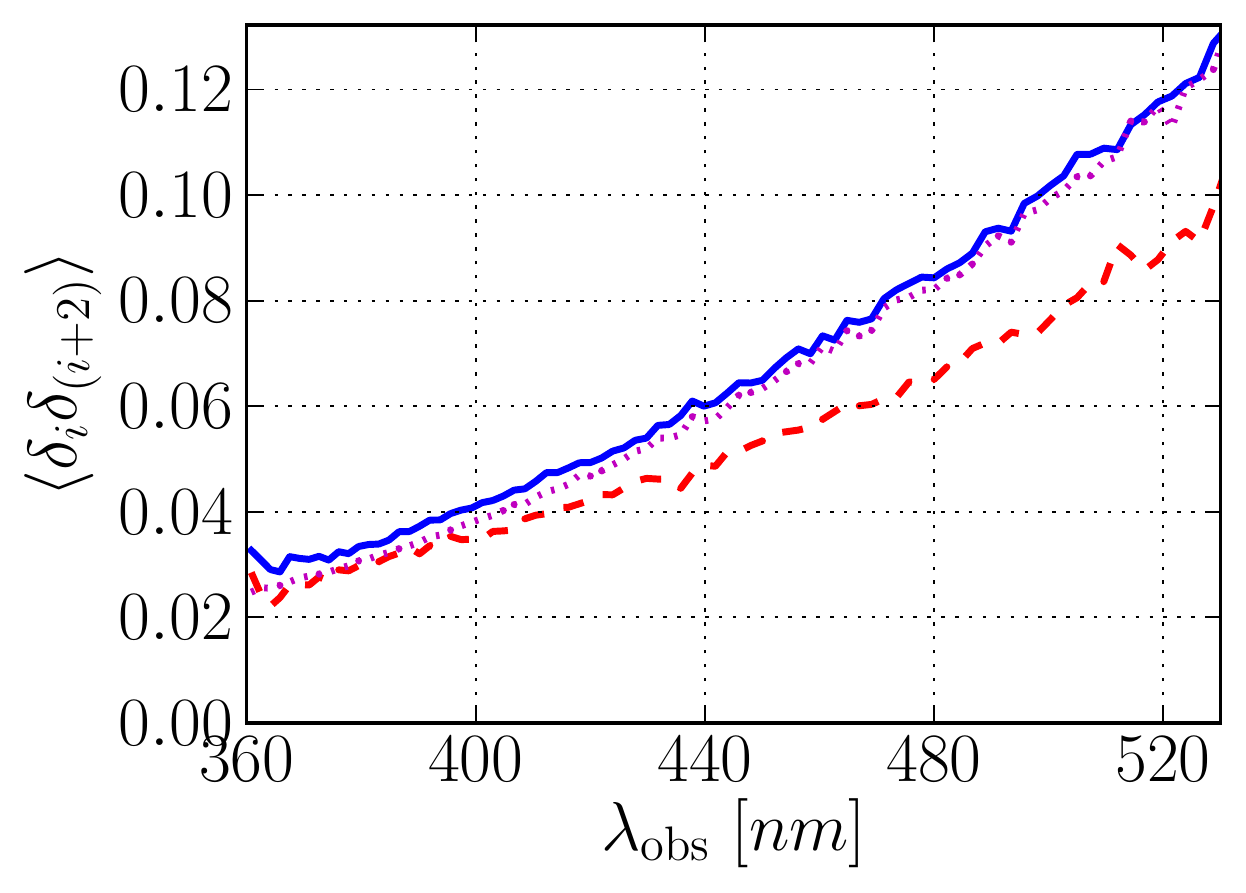}
 \end{center}
 \caption{ 
 Correlation $\xioned$ between a pixel and its nearest neighbor  
 ($\log_{10}(\lambda_1/\lambda_2)=10^{-4}$, left) and
 next-to-nearest neighbor 
 ($\log_{10}(\lambda_1/\lambda_2)=2\times10^{-4}$,right). 
 The data are the red dashed lines, 
 the expanded mocks blue solid line, 
and raw (noiseless) mocks  the magenta dotted line. 
}
 \label{fig:mock_data_cov1d}
\end{figure}

We first discuss $\xioned$ for small separations.
Figure \ref{fig:mock_data_cov1d} compares $\xioned$ 
for mocks and data for
the first two non-zero separation bins: 
$\log_{10}(\lambda_1/\lambda_2)=3\cdot10^{-4}$ (left) and 
$\log_{10}(\lambda_1/\lambda_2)=6\cdot10^{-4}$ (right). 
In these cases, the blue line (standard
mocks) and the magenta line (noiseless mocks) closely follow one another. This
is because our model assumes that the noise in different mock pixels 
is uncorrelated (this is not true for real data). 
The disagreement between mocks and data is at the
15-20\% level and could be due either to an inaccurate input correlation
function on small scales, or to inaccurate modeling of the instrument. 
On the clustering side, the small scales of these mocks are not supposed 
to be accurately modeled (due to resolution limitations), 
since the main goal is the large scales and BAO measurement. 
On the instrumental side, we do not simulate the coaddition process of individual exposures, 
that introduces correlations among nearest neighboring pixels in real data.

%\af{I have several objections on the sentences above: i) "This is because
%our model assumes taht the noise in different pixels is uncorrelated"? Are we
%sure about this? I would expect that the level of correlated noise was quite 
%small... One could test this by measuring the correlation function for 
%different sub-samples split by SNR, or looking at the correlated noise from
%different exposures, but I'm quite sure this is not the main issue here.
%I'd rather say that our mocks are not aimed at describing the smallest scales.
%By the way, are the "noiseless mocks" continuum fitted, or you use the right 
%continuum? Because continuum fitting would also distort (quite strongly) 
%the 1D correlation function... Also, do they include HCDs/metals? 
%ii) why would UV fluctuations be important at these particular 
%scales, and not on larger scales? Do we have any explanation for this? I would
%expect the effect of UV fluctuations to be important on much larger separations.
%iii) I agree that resolution / pixel width effect might be important, together 
%with the effect of binning in our skewers (the cells are too large).}

The correlations for larger pixels separations are shown in figure \ref{fig:mock_data_xi1d} for three redshift
bins.
Outside
the SiIII bump at $\log_{10}(\lambda_1/\lambda_2)\approx32\times10^{-4}$, 
the mocks clearly have less correlation than
the data.  On the bump, the mocks appear to have a much stronger
redshift dependence of the absorption.

The disagreement between the data and mock $\xioned$ means that the covariance matrices of the two will be slightly different
(section \ref{sec:covariance}).
Given the uncertainties in BAO measurements are dominated
by pixel variance and the covariance of neighboring pixels, we will leave these issues to future work.

An alternative way to study correlations within individual spectra is
the Fourier transform of $\xioned$, the one-dimensional power spectrum,
$\pkoned$, a very powerful probe to constraint cosmological models
\cite{2006ApJS..163...80M, 2013A&A...559A..85P}.
However, we do not expect that
the mocks spectra presented here will give a realistic spectrum for all $k$,
especially at small scales where our treatment is rather crude (see section
~\ref{sec:absorption}). We have also assumed a constant resolution over the whole mock forests.

We computed the 1D power spectrum following the same procedure as in 
\cite{2013A&A...559A..85P} for a sub-sample of 16,000 mock spectra having 
signal-to-noise ratio greater than 2 and mean resolution below 85 $\kms$ (same 
cuts as for real data). 
The flux fluctuations $\delta_F$ were obtained by dividing out the mean stacked 
flux over forests. 
We used a Fast Fourier Transform to convert the fluctuations to $k$-space to 
estimate the power spectrum. 
The noise contribution to the power was subtracted, assuming it is constant 
over $k$. 
The resolution and pixelization kernel (Eq.~\ref{eq:kernel}) was also divided 
out. 
%\textbf{
This measurement was done in four redshift bins equally divided between 
$2.1< z < 2.9$. 
We analyzed mock spectra with no noise added (only the continuum multiplied by 
the absorption field) in order to see effects of binning and resolution on the 
power. 

The analysis on full expanded spectra needs an estimator for the noise 
mis-estimates. 
In \cite{2013A&A...559A..85P}, this estimate was performed using two different 
procedures, the first uses individual exposures of each spectra, the second 
uses flat spectral regions redwards of the \lya\ emission peak. 
None of these procedures can be applied on mock spectra since individual 
exposures are not produced, and no quasar with $z<2.15$ is created meaning that 
no region of mock spectrum redwards of \lya\ falls into the blue end of the 
spectrograph, which would allows us to estimate the true noise. 
Therefore, the 1D power spectrum measurement on expanded spectra was performed 
using the true pixel noise, which is constant in $k$-space.

\begin{figure}[t]
 \begin{center}
  \includegraphics[width=0.6\textwidth]{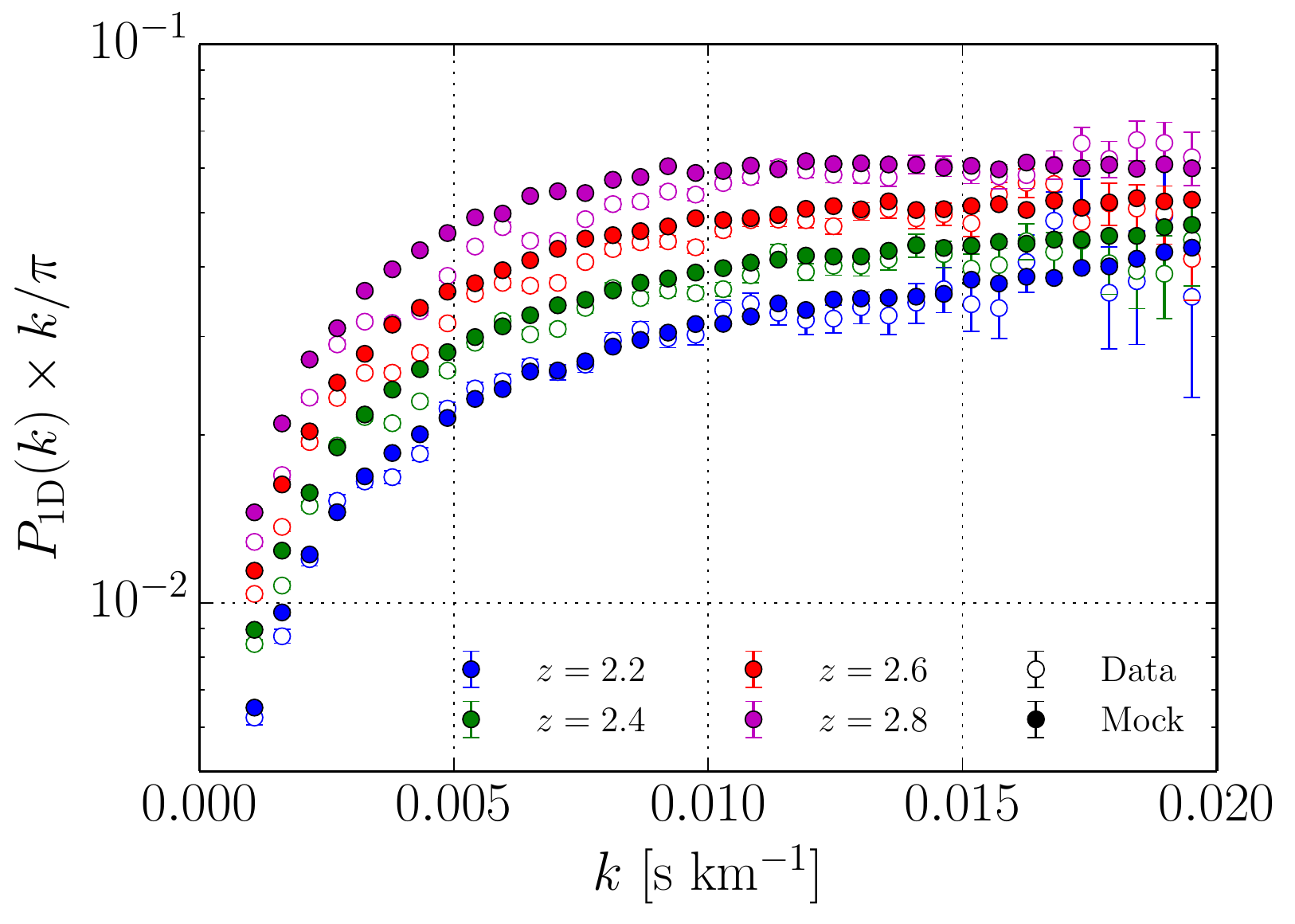}
 \end{center}
 \caption{Line-of-sight power spectrum computed on 16,000 mock spectra (filled 
 circles) compared to the measurement on DR9 forests (open circles) on 4 
 redshift ranges.
%\jr{suggest leaving out the two highest redshift bins--they do not
%really concern us here and the figure is too cluttered as is}
}
 \label{fig:pk1d_mock}
\end{figure}
Fig.~\ref{fig:pk1d_mock} shows the estimated 1D power spectrum of mock spectra, 
compared with the measurement performed on DR9 data \cite{2013A&A...559A..85P}. 
The overall shape of power spectra is in good agreement with data.
% on large scales,  $k < 0.015$~s~km$^{-1}$. 
The power spectrum on the lowest redshift bin, $z=2.2$, shows the best 
agreement with data while the higher redshift measurements have up to 20\% 
more power than data on large scales ($k < 0.015$~s~km$^{-1}$). 
This increase in power at high redshift is related to the increase in pixel 
variance observed in Fig.~\ref{fig:mock_data_var1d}.
%\textbf{
On small scales ($k > 0.015$~s~km$^{-1}$), data and mocks show an agreement within 1$\sigma$ 
for most points, but this is due mainly to the large errors on data points. 
As we saw in the $\xioned$, the clustering on small scales is not correctly modeled in our mocks.

We remind that covariance matrix measurements for BAO in data are not derived from mocks, but from the data itself. Therefore, discrepancies seen here in the small scale clustering does not have an important role in the final measurement.

%\af{I find puzzeling to say that $\xioned$ doesn't look good but $\pkoned$ 
%does. I'd stick to the explanation that there are small differences in the
%small scale clustering, and that this will result in differences in the 
%covariance matrices. And remind the reader that the covariance matrix for the
%data was not derived from the mocks (otherwise it would be bad).}

\subsection{Metals}
\label{sec:test_metals}

\begin{figure}[t]
 \begin{center}
  \includegraphics[width=\textwidth]{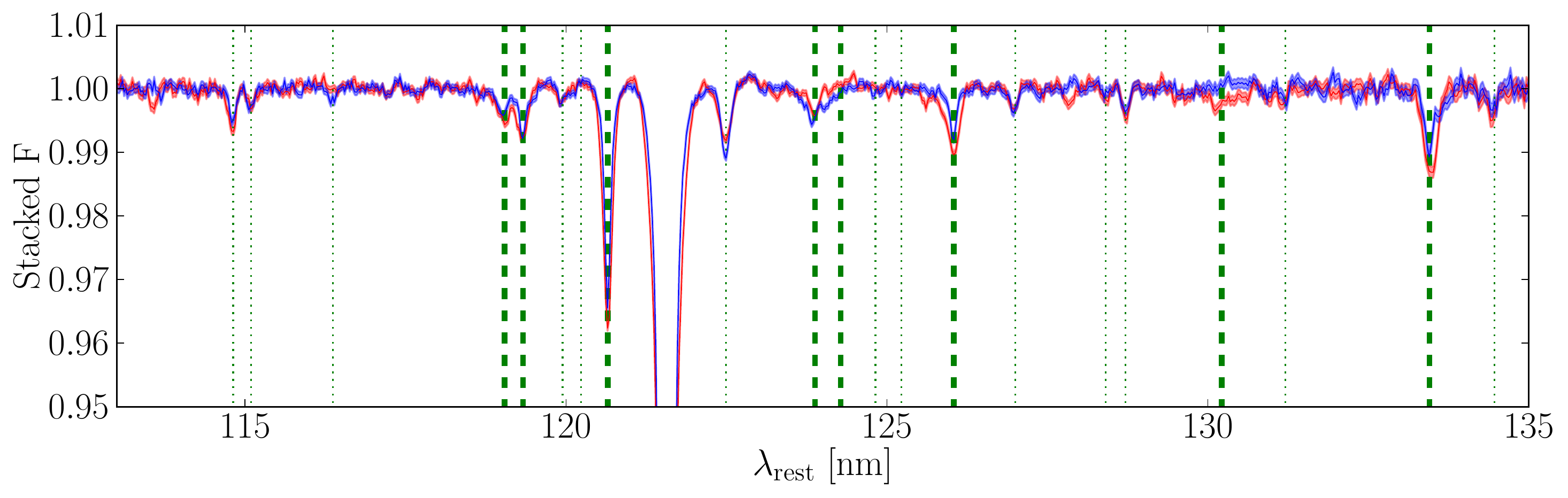}
 \end{center}
 \caption{Stack of \lya\ absorption lines in their rest-frame showing the metal 
 absorption for a mock realization (blue) and for data (red). The 1$\sigma$ bootstrap errors are represented by the color shaded regions (same as lines). 
 Dashed thick green lines show the position of the implemented metals correlated 
 with \lya\ absorption, while thin doted green lines are ``shadows'' (metal-metal 
 correlations).}
 \label{fig:metalstack}
\end{figure}

Metal absorption correlated with \lya\ absorption was included in mock forests 
following prescription described in section~\ref{sec:exp_Metals}. 
In order to test our method of introducing metal absorption, we stacked \lya\ 
absorption in our mock forests following a modified method from 
\cite{2014MNRAS.441.1718P}.
In Fig.~\ref{fig:metalstack} the resulting stack for \lya\ transmission values 
in the range $0.05 < F < 0.15$ is shown as a function of rest-frame wavelength, 
in comparison with the stack using real data. 
Errors were computed by bootstrap. 
%\jr{What error bars? They aren't
%on the figure.  Are they relevant?  should we give details}. 
The agreement for most of the lines is at the sub-percent level, in particular 
for the stronger ones. 
Similar agreement is also observed in stacks of absorbers with different 
transmission values ($F< 0.05$ and $0.15 < F < 0.4$).
This validates our implementation, constructed with the purpose of 
investigating systematic effects of metals on the BAO measurement. 
The effect of these metals on the correlation function is discussed in 
section~\ref{sec:3d_clustering}.

\section{Three-dimensional statistical properties of the mock and real spectra}
\label{sec:covsec}

In this section we compare the 3D correlation function of the
mock and real spectra and their covariances.
We do this first for the mock spectra without high-column-density
and metallic absorbers.
We then  
study the effects of 
of these two absorbers on the measured
correlation function.

\subsection{Three dimensional correlation function}
\label{sec:3d_clustering}

%In this section we compute the 3D correlation function of mock catalogs and we
%compare results with the same measurement on data. 
%We also compare the covariance matrices of mocks to those of data. 
%As an application, we also study the effects of adding metals and 
%Lyman-limit systems to the forest.

%\af{Actually, the effect of LLS on the covariance matrix is discussed in 
%the next sub-section. The effect of metals on the correlation function is 
%also discussed in the next sub-section, what doesn't make much sense.}

%\af{In the previous section, we describe how to add HCD systems, including 
%both LLS and DLAs. Here, we only talk about LLS, so we should explain why.}

%For a given absorption field (see section \ref{sec:absorption}) the user has
%the choice of adding metals and/or Lyman-limit systems to the \lyaf\ of the 
%final fully expanded mocks. We start with the simplest ``baseline'' mocks
%that do not include these features and compare the 3D correlation function
%of mocks and data.%

%Since the format of our expanded mocks is the same as the BOSS \emph{per-object}
%file (\emph{REFERENCE NEEDED}), it is straight-forward to apply the methods
%from \cite{delubac_baryon_2015} but using mocks as input. 

\begin{figure}[t]
 \begin{center}
  \includegraphics[width=0.45\textwidth]{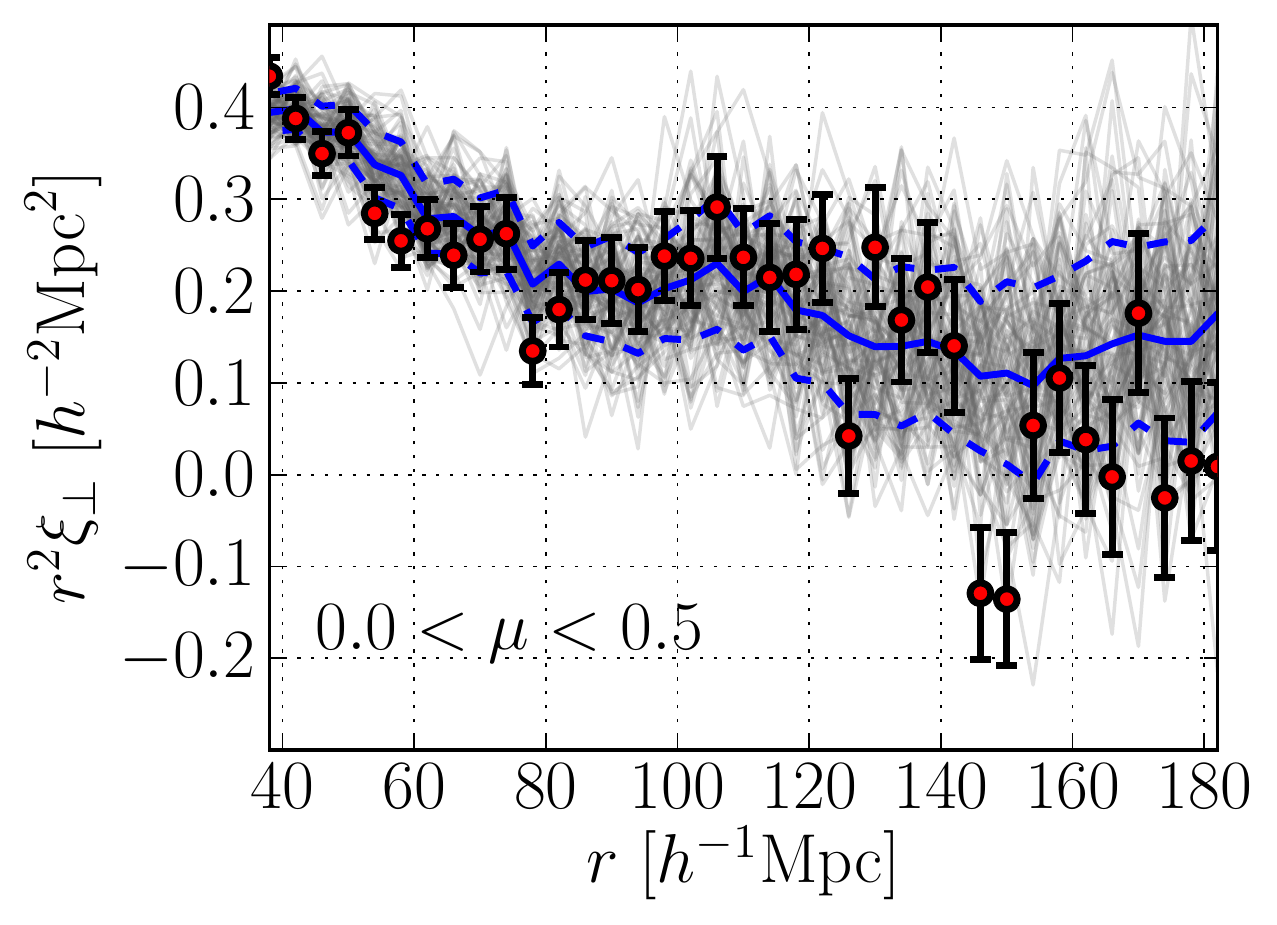}
    \includegraphics[width=0.45\textwidth]{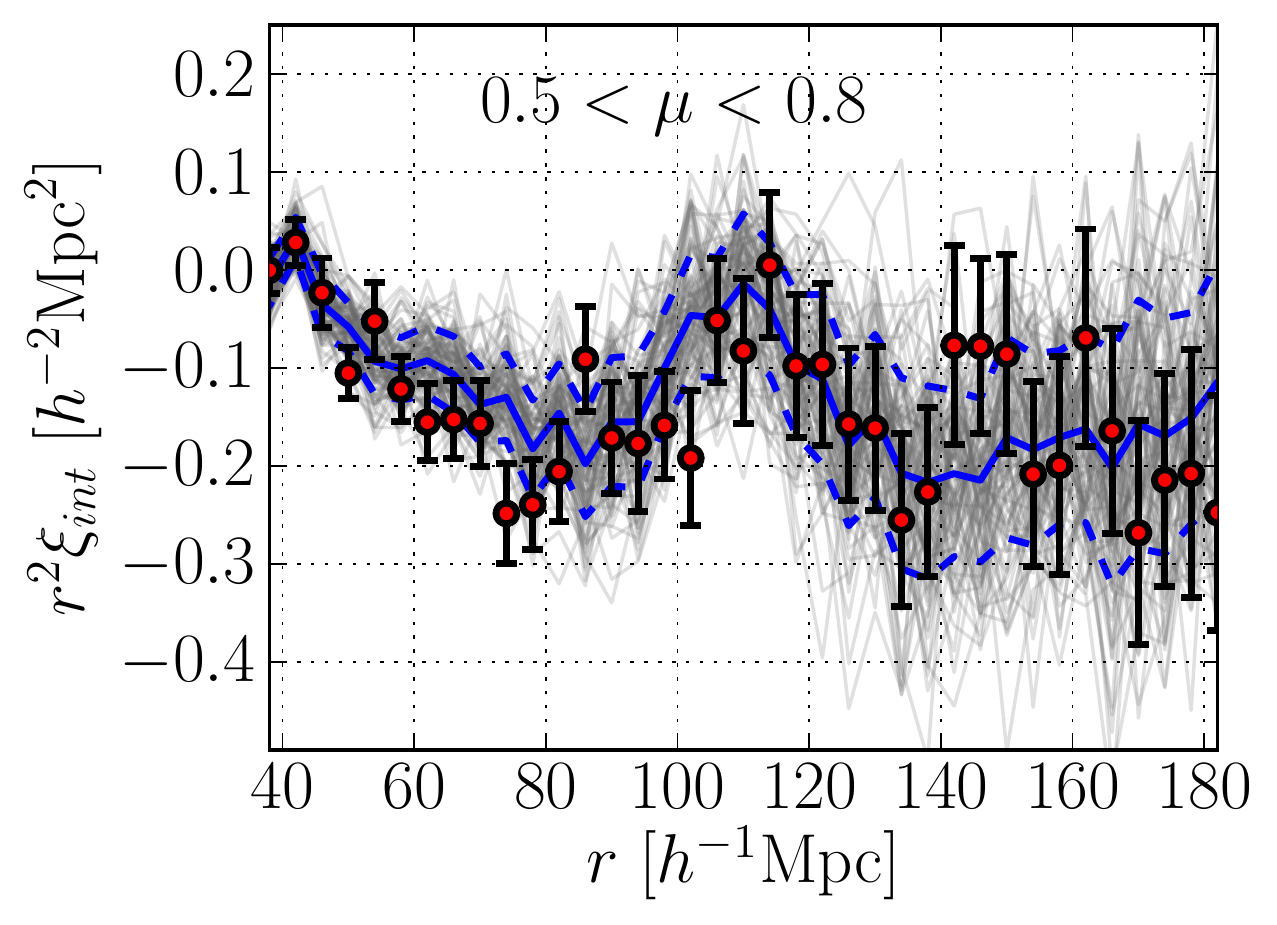}
     \includegraphics[width=0.45\textwidth]{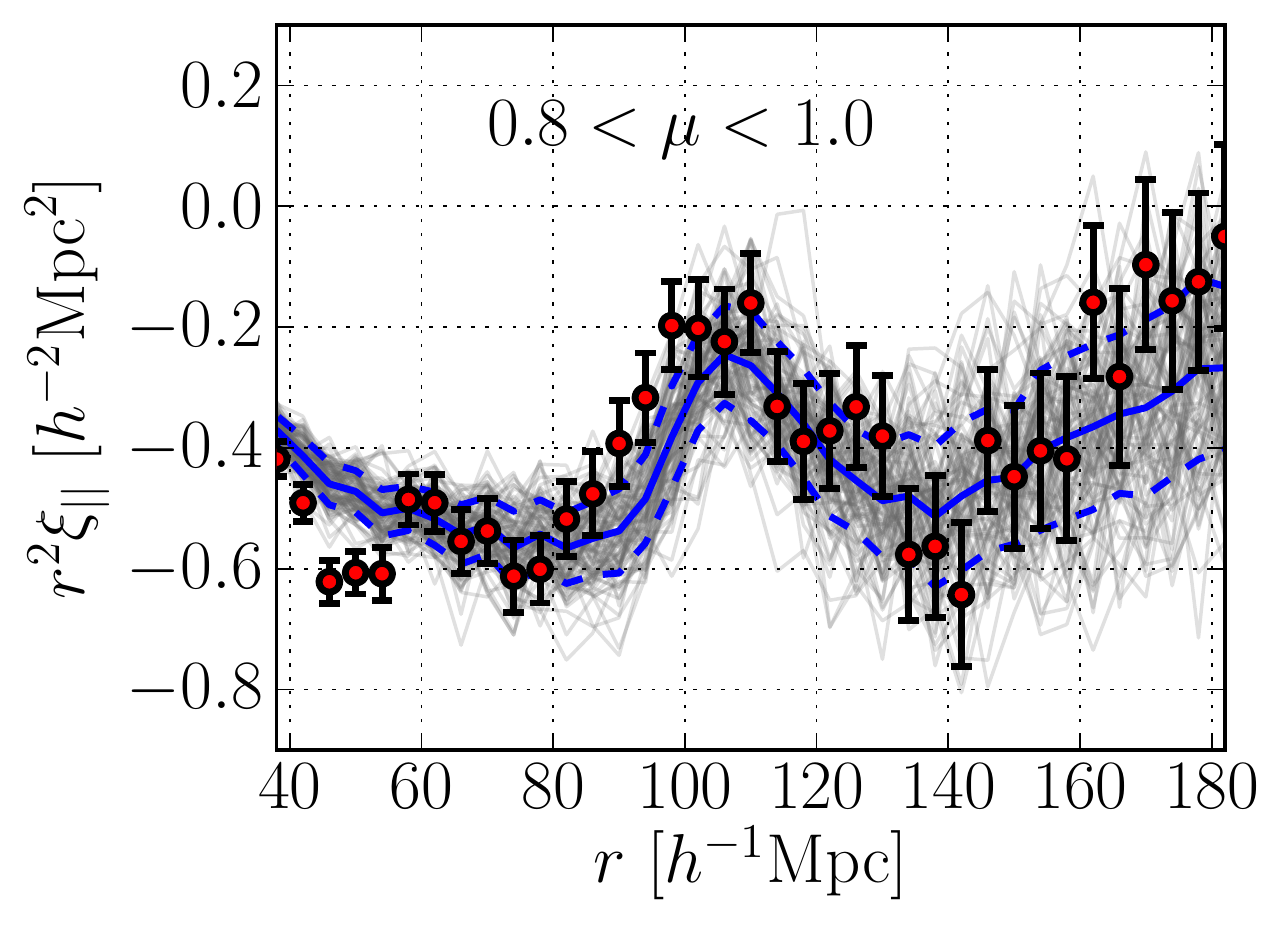}
 \end{center}
 \caption{ 
 Measured \lyaf\ 3D correlation functions represented by averages over 
 $\mu = \rpar/r$ ranges, useful to see the effect of redshift-space distortions (top left: $0<\mu<0.5$, top right: $0.5<\mu<0.8$, 
 bottom: $0.8< \mu < 1.0$). 
 Red points and their error bars show the data measurement whereas gray lines 
 show the same for the 100 realizations of DR11. 
 The blue solid lines show the mock average and blue dashed lines their 
 $\pm 1\sigma$ standard deviation around the mean.}
 \label{fig:mock_data_wedges_scatter}
\end{figure}

The correlation function
is measured using eqn. \ref{eq:2ptcorrelation}
as a function of the transverse separation
($\rperp$) and the parallel separation ($\rpar$) between pixels. We 
organize these data in bins or $r\equiv\sqrt{\rperp^2+\rpar^2}$ 
and $\mu\equiv \rpar/r=\cos\theta$ (where $\theta$ 
is the angle between the pixel-separation vector and the line of sight).
We finally consider three ``wedges'' in $\mu$: $0<\mu<0.5$, $0.5<\mu<0.8$ and 
$0.8<\mu<1$ and compute the average of the correlation function in each wedge 
as a function of $r$ to obtain $\xi_\perp$, $\xi_{int}$ and $\xi_\parallel$
respectively. The three panels in figure 
\ref{fig:mock_data_wedges_scatter} show the results. The gray lines
correspond to individual mock realizations and illustrate the
spread. The blue lines show the mean (solid) and the $1\sigma$ limits (dashed)
for the sample of mocks. The red points with error bars are the results
from the data
\cite{delubac_baryon_2015}.

The agreement between the mocks and data is good. The most noticeable
difference is a mismatch in the position of the BAO peak in $\xi_\parallel$.
This difference is characterized in \cite{delubac_baryon_2015} as
a $\approx1.5\sigma$ discrepancy between the predicted position of the BAO
peak by the fiducial cosmology of the mocks and that measured from data.

\subsection{Covariance.}
\label{sec:covariance}

The covariance matrix of the 3D correlation function bins can be estimated 
directly from the data using the subsampling technique developed in 
\cite{delubac_baryon_2015}.
This internally estimated covariance for a given mock can be compared
with the variability among the 100 mock realizations characterized
by its variance:
\begin{equation}
Var\left[\xi_{A}\right]
 = {1\over N_{mocks}-1}\sum_{i=1}^{N_{mocks}}(\xi_A-\bar{\xi}_A)^2
\label{eq:var_mocks}
\end{equation}
where $N_{mocks}$ is the number of mocks (100) $\xi_A$ is the 3D correlation
function measured at a separation bin $A$ and $\bar{\xi}_A$ is the average
correlation function in the mocks sample.
Since
non-diagonal correlations are too small to estimate with only 100 mocks, we only
compare the variances.

\begin{figure}[t]
 \begin{center}
 \includegraphics[width=0.49\textwidth]{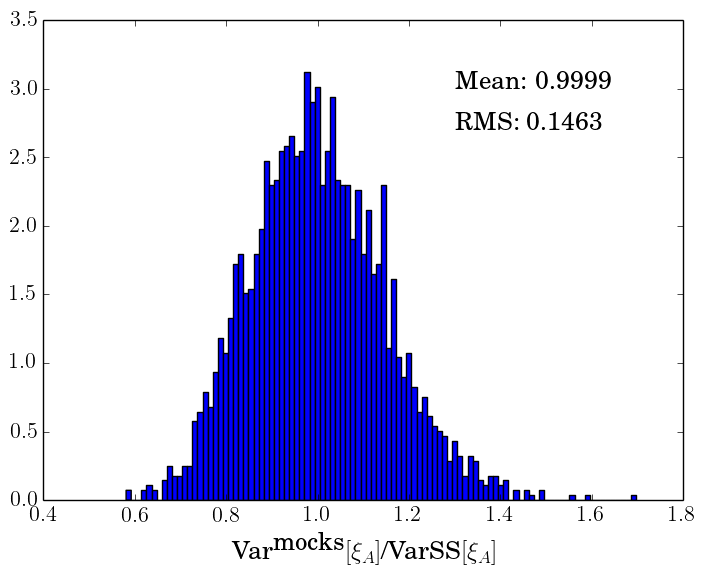}
 \end{center}
 \caption{
 Histogram of the ratio of the variance of $\xi_A$ calculated from the 
 fluctuations in the 100 mock and using the sub-sampling method. Each
 histogram entry corresponds to a bin in separation.
}
 \label{fig:mock_mock_variance}
\end{figure}

Figure \ref{fig:mock_mock_variance} shows a histogram of the ratio of  
variances estimated using equation \ref{eq:var_mocks} and that obtained
using the subsampling method from \cite{delubac_baryon_2015}. The agreement
between both calculations of the variance is excellent. This validates our methods to compute the covariance matrix, showing the importance of this set of mock catalogs for the measurement on real data.  

%\af{I think this is the most important figure in the paper. It shows that 
%thanks to the mocks we have shown that our method to estimate the covariance 
%matrix is valid. I would try to give it more weight, both here and in the
%abstract / conclusions.}

\begin{figure}[t]
 \begin{center}
 \includegraphics[width=0.49\textwidth]{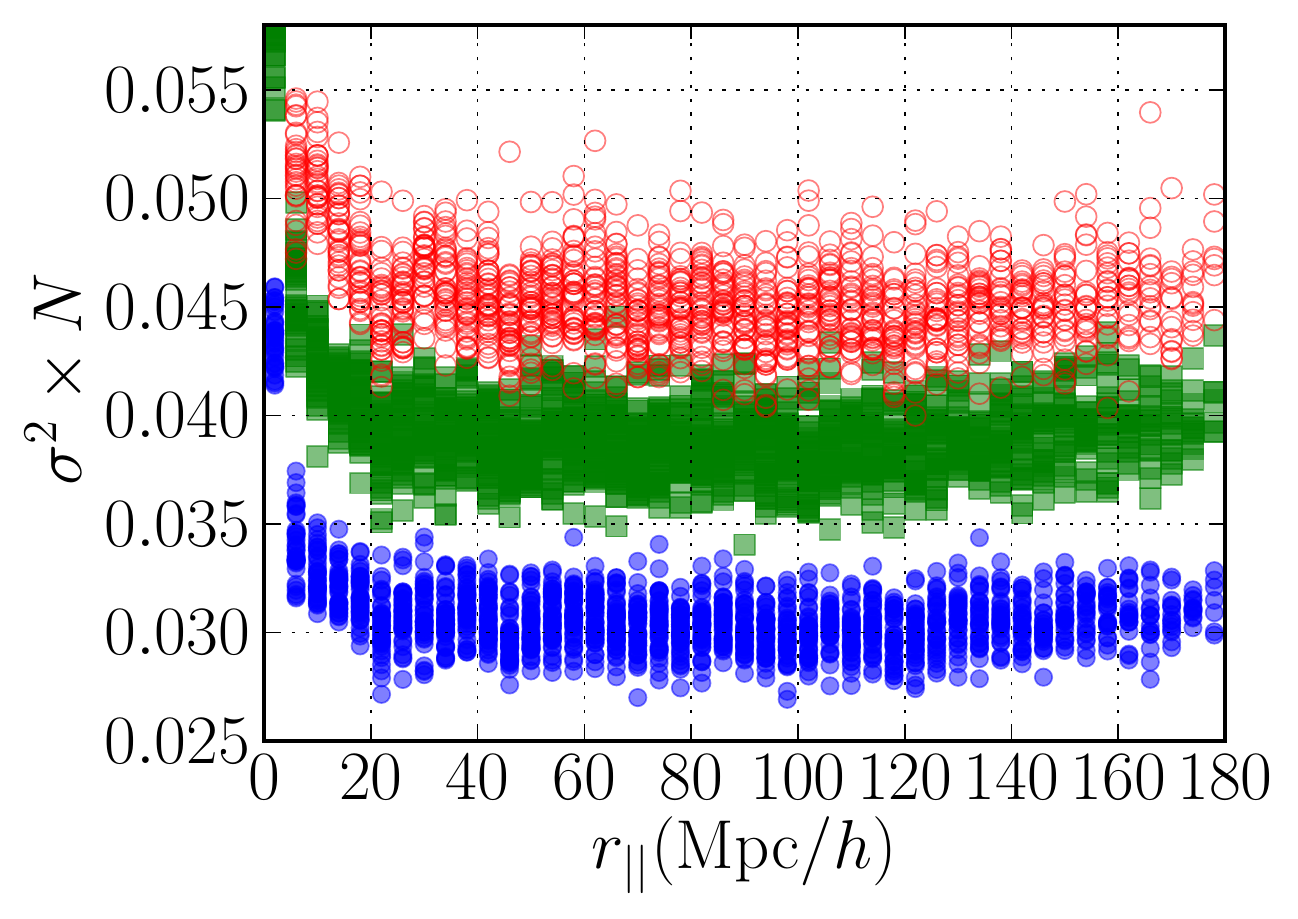}
 \end{center}
 \caption{Variance $\sigma^2 = Var[\xi(\rpar,\rperp)]$ of the 3D 
 correlation function times the number of pairs $N$ as a function of parallel 
 separation $r_{||}$. 
 Red empty circles show data, blue filled circles (resp. green squares) show a 
 single mock realization measurement without (resp. with) Lyman Limit Systems.
 Remark that the error bars of mock containing LLS correspond to a more biased 
 correlation function (Fig.~\ref{fig:DLAMet_wedges}).}
 \label{fig:mock_data_variance}
\end{figure}

Next we compare the variances in mocks and data. Since the number of quasars
in the mocks is not exactly the same as in real data, we compare the product
of the variance and the number of pairs for each bin in separation. 
Figure \ref{fig:mock_data_variance} shows the product of the variance and
the number of pairs as a function of transverse separation for data (in red)
and mocks (in blue). This figure shows that the mocks have 30\% less variance
than the data. This difference may come from the differences in the 
one-dimensional correlation function (see figure \ref{fig:mock_data_var1d}).

\begin{figure}[t]
 \begin{center}
 \includegraphics[width=0.45\textwidth]{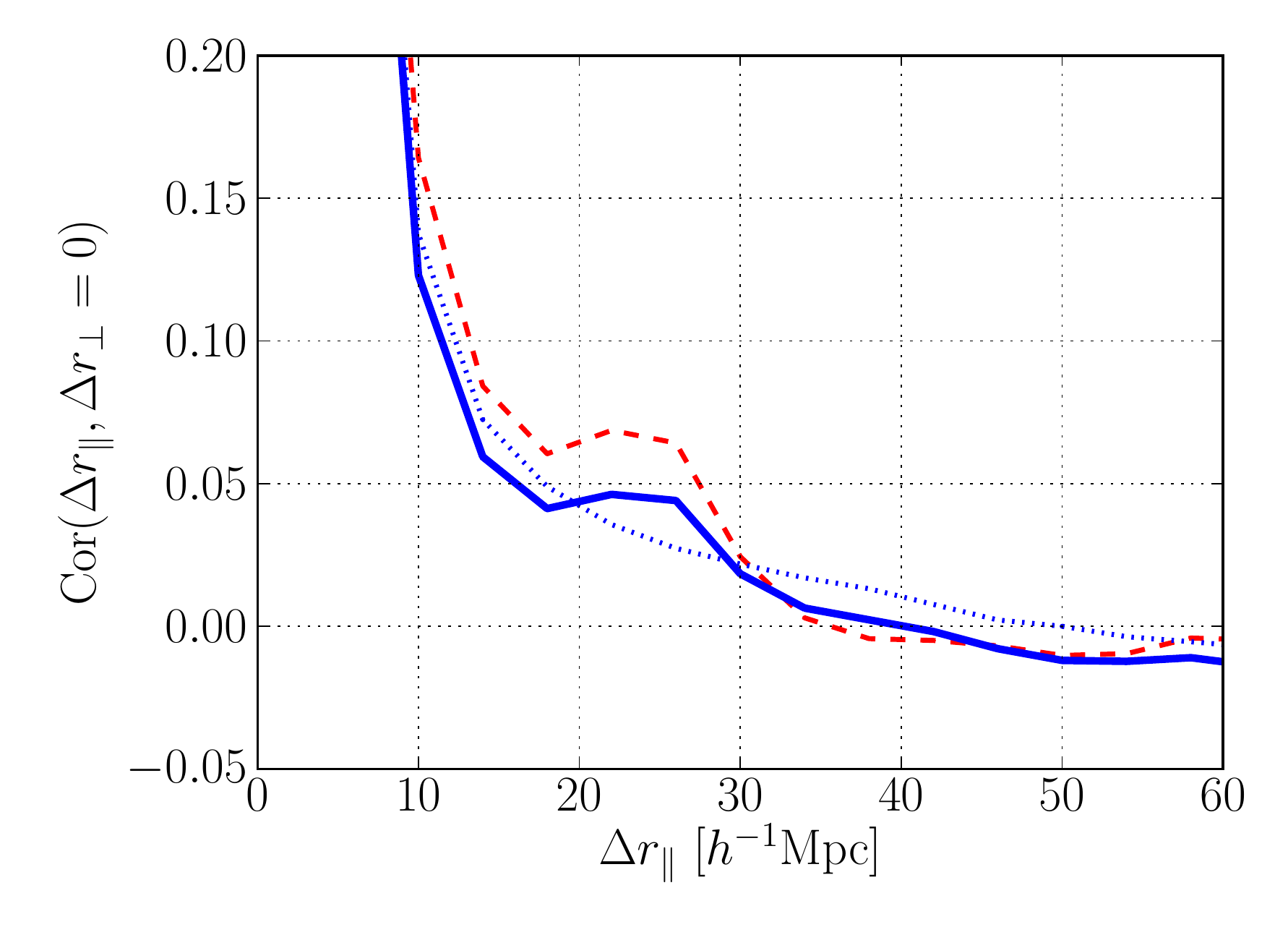}
 \includegraphics[width=0.45\textwidth]{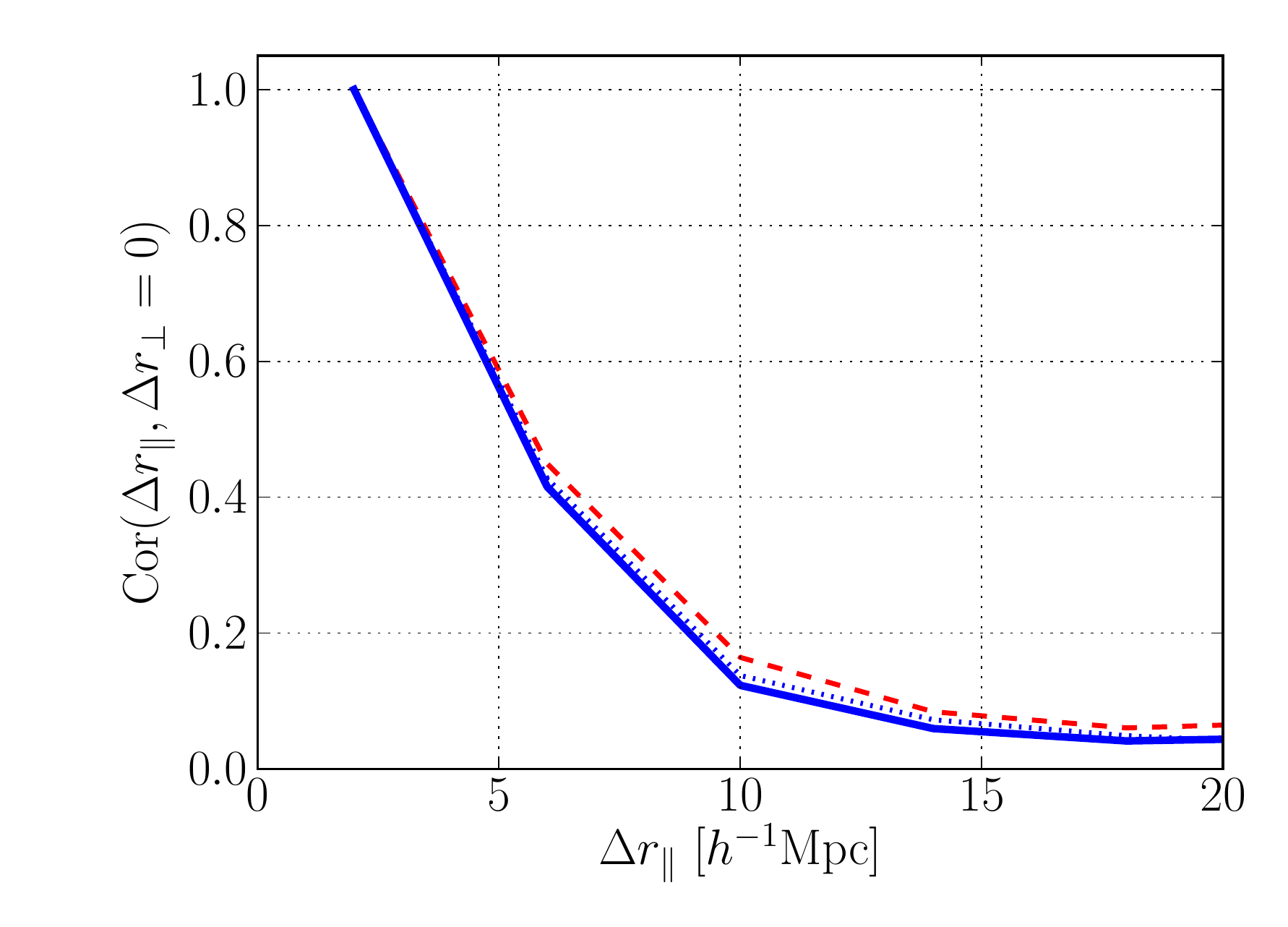}
 \includegraphics[width=0.45\textwidth]{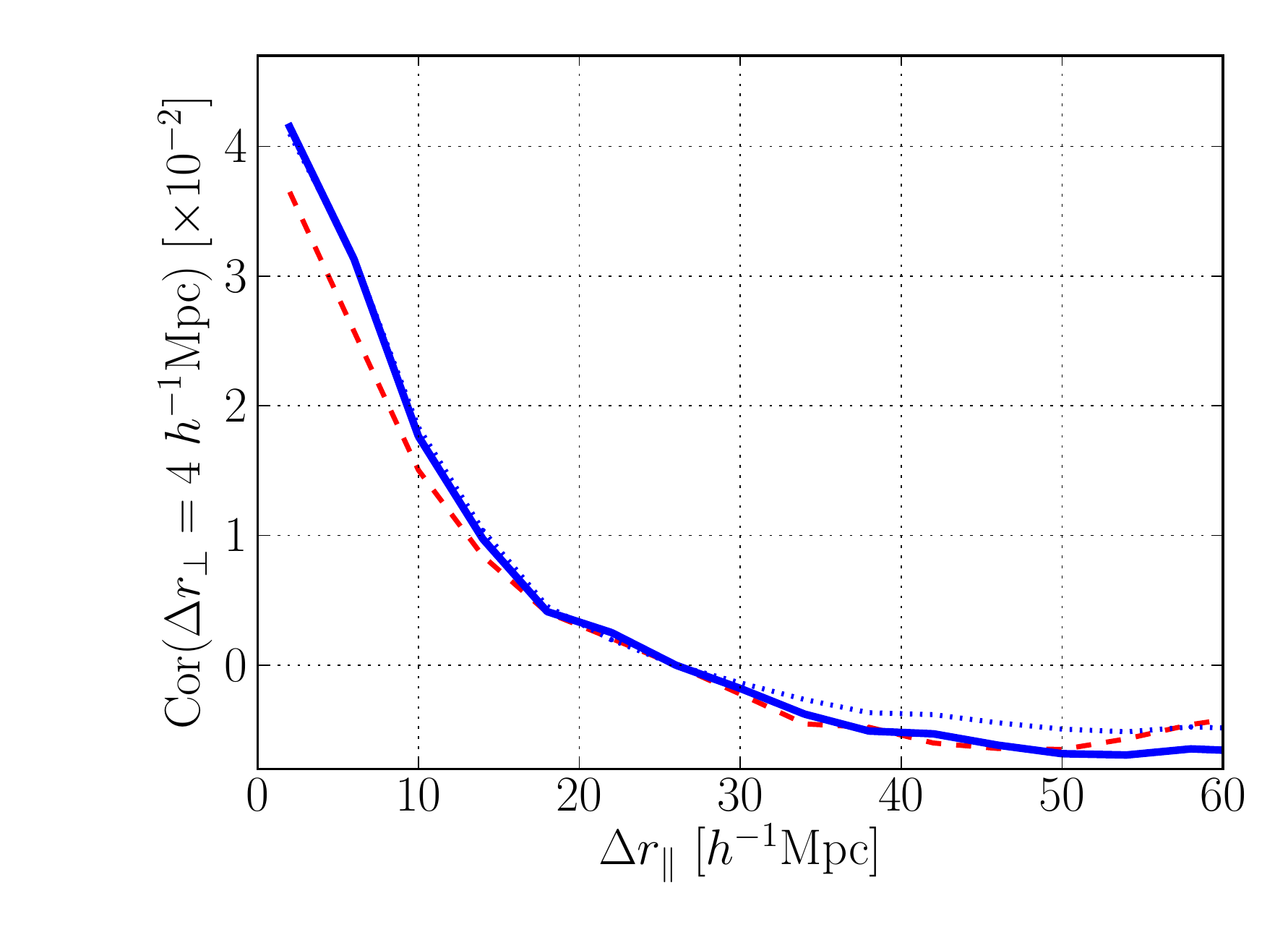}
 \includegraphics[width=0.45\textwidth]{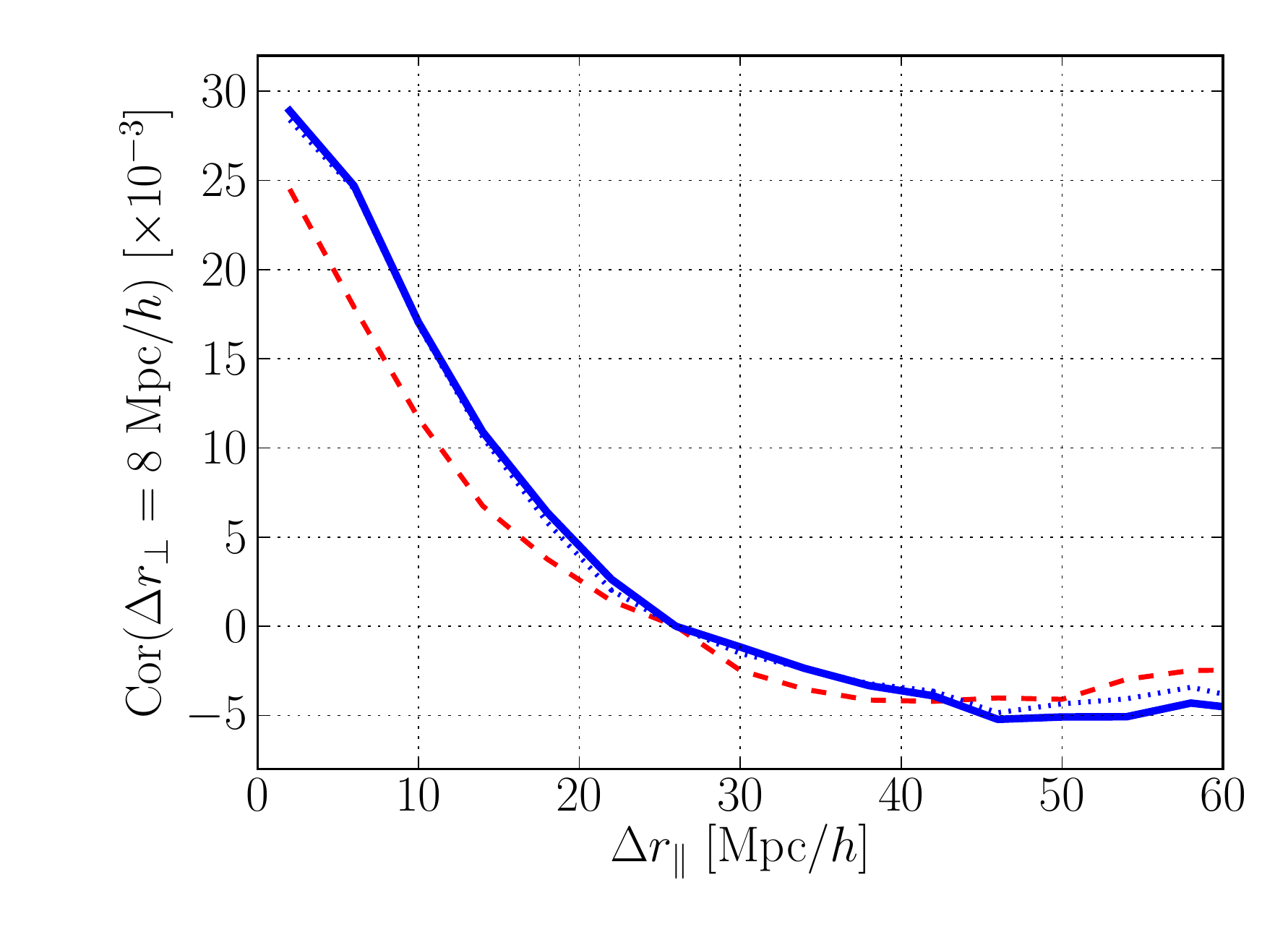}
 \end{center}
 \caption{
%\jr{We need a 4th figure that gives the low deltarpar bins for deltart=0}
Correlation coefficient 
 $C(\rperp,r^\prime_\perp,\rpar,r^\prime_\parallel) 
  		/ \sqrt{Var(\rperp,\rpar)Var(r^\prime_\perp,r^\prime_\parallel)}$ 
 of the 3D correlation function as a function of parallel separation difference 
 $\Delta \rpar = \rpar - r^\prime_\parallel$. 
 The values shown are the average of all correlation matrix elements with the 
 same $\Delta \rpar$ and $\Delta \rperp$, for $\Delta \rperp = 0$ 
 (top left and top right), $4 \hMpc$ (bottom left) and $8 \hMpc$ (bottom right panel). 
 Data is shown in red dashed lines, mock measurements in blue, with (solid) 
 and without metals (dotted).}
 \label{fig:mock_data_covariance}
\end{figure}

We also compared non-diagonal terms of the covariance matrix. 
Following \cite{delubac_baryon_2015}, we assume that this matrix is a function of 
$\Delta \rpar = \rpar - r^\prime_\parallel$ and 
$\Delta \rperp = \rperp - r^\prime_\perp$, and thus 
compute of the average of all 
correlation coefficients (covariance normalized by the geometric mean of 
variances) with same $\Delta \rpar$ and $\Delta \rperp$.
In figure \ref{fig:mock_data_covariance}, we plot these averaged correlations 
as a function of $\Delta \rpar$ for the three first transverse bins, 
$\Delta \rperp = 0, 4$ and 8 $\hMpc$. 
In the first transverse bin, where the covariance matrix is determined by the 
line-of-sight correlation function, we clearly see a peak at 
$\rpar \sim 25 \hMpc$ corresponding to \lya-SiIII correlations. 
This peak is, as expected, absent from the covariance matrix 
of metal-less mocks. 
As before, data shows slightly more correlation on small scales. 
The covariance falls quickly to zero as $\Delta \rperp$ increases. 
For these bins, data and mock non-diagonal terms of the covariance matrix 
generally agree in shape but data shows smaller correlation coefficients 
likely due to the differences in the one-dimensional correlation function.

\subsubsection{Effect of Lyman-limit systems.}

\begin{figure}[t]
 \begin{center}
 \includegraphics[width=0.45\textwidth]{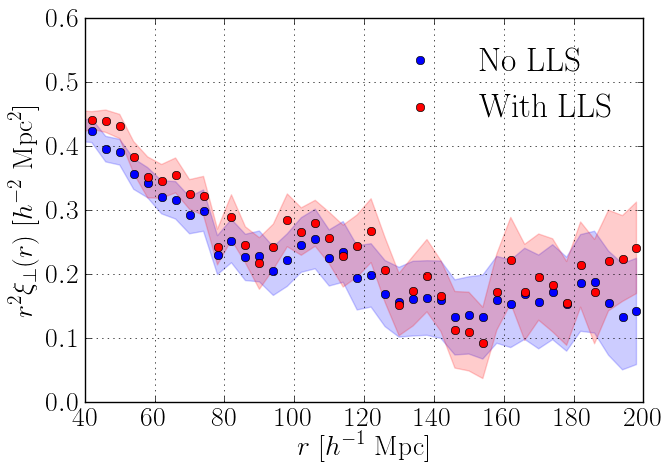}
 \includegraphics[width=0.45\textwidth]{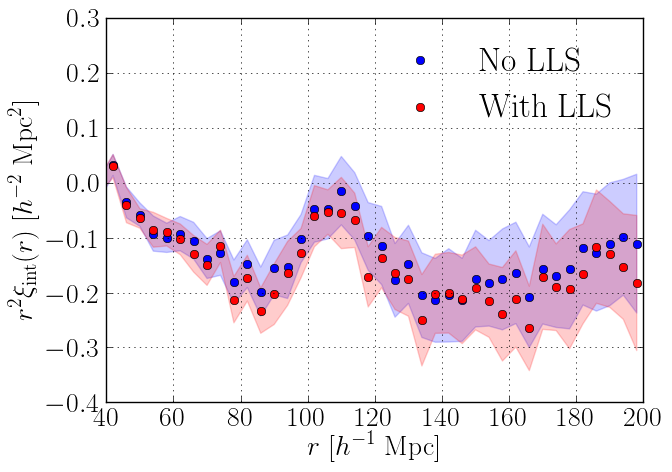}
 \includegraphics[width=0.45\textwidth]{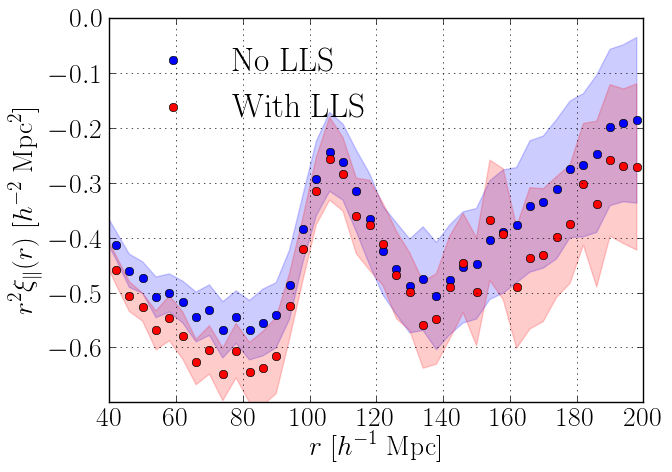}
 \end{center}
 \caption{Effect of Lyman Limit Systems (LLS) on the 3D correlation function 
 represented by three wedges $0<\mu<0.5$ (left), $0.5<\mu<0.8$ (center) and 
 $0.8 < \mu < 1.0$ (right panel). These are averages over 100 (resp. 10) realizations containing (resp. without) LLS. Colored shaded regions show the 1$\sigma$ scatter around the mean.
} 
 \label{fig:DLAMet_wedges}
\end{figure}

Thus far, the comparison has been performed using mock catalogs with no
high-column density absorbers 
even though they are present in data.
In this section we see the effect of adding Lyman-limit systems (LLS) to
the spectra. Since DLAs can be identified in the data and treated
separately, 
we excluded from our analyses all forests containing DLAs ($N_{\rm HI} > 10^{20.3}\cm^2$), while conserving forests with LLS ($10^{17} < N_{\rm HI} < 10^{20.3} \cm^2$).
In Fig.~\ref{fig:DLAMet_wedges} the 
stacked 3D correlation function of mocks 
containing LLS is compared with the stack of 
realizations without them. 
As already observed by \cite{2012JCAP...07..028F}, the inclusion of high column 
density systems increases the bias of the correlation function. 
This is now shown also taking into account the BOSS spectrograph properties 
and using the standard BAO \lya\ analysis. 
This effect comes from the increased absorption from LLS that reduces the mean 
transmission $\bar{F}$ and also increases the intrinsic variance 
$\sigma^2_\mathrm{LSS}$ of the absorption field. 
The increase in bias of the correlation function depends on two factors. 
First it depends on the number density of these systems and the dependence of 
this number with redshift. 
Second, it depends on how these systems are related to the underlying density 
field. On real data, the distribution of Lyman limit systems is not known to 
high precision \cite{2011ApJ...743...82M}, therefore the used input mock LLS 
distribution might not be accurate. 
Furthermore, due to our method of including LLS on forests (see section
~\ref{sec:DLAs}), the redshift distortion parameter for LLS, $\beta_{\rm LLS}$, 
is the same as for the \lyaf, $\beta = 1.4$, which is also an approximation 
(we expect that $\beta_{\rm LSS} < \beta$). 
Therefore quantitative results of their effects on the correlation function 
are not reliable, even though their qualitative effect is 
correct. 
It is important to remark in Fig.~\ref{fig:DLAMet_wedges} that these systems 
do not change the position of the BAO peak or its width. 
However, measuring bias or redshift space distortions need to carefully take
into account 
the additional clustering coming from Lyman limit systems.

\subsubsection{Effect of metals.}

\begin{figure}[t]
\centering
\includegraphics[width=0.5\textwidth]{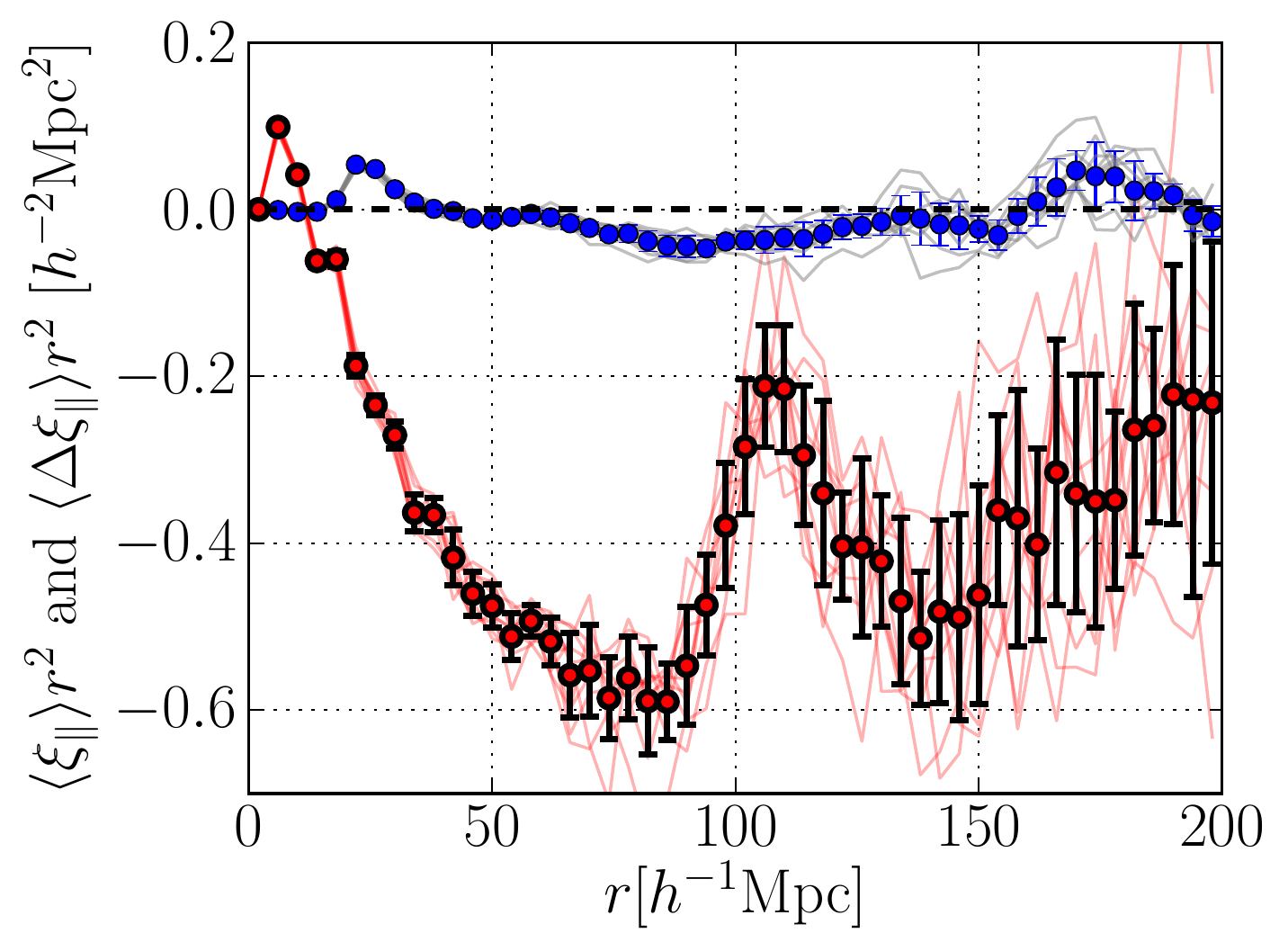}
\caption{The effect of metals on the stacked measurement correlation function of 
10 mock sets. The difference between metal and metal-less mock correlation function averaged over $0.8<\mu<1.0$ (blue points) compared with the measurement itself (red points).
The light red and blue lines show the results for individual mock sets. Error bars are the standard deviation of the 10
estimates.}
\label{fig:metal_effect}
\end{figure}

In addition to Lyman limit systems, real forests also contain metal absorption 
that 
cannot  be identified due to the low signal-to-noise of spectra 
or confusion between source transitions.
We included metal absorption in mock spectra (see section~\ref{sec:exp_Metals}) 
and we found that their effect on the correlation function and its errors is 
smaller than the effect of LLS. 
The main effect of metal absorption is seen on correlations of pixels with 
separations nearly aligned to the line-of-sight ($\mu \sim 1.0$), on bins with 
small transverse separation ($\rperp \sim 0$). 
Figure~\ref{fig:metal_effect} shows the difference between correlations of metal and metal-less mocks (blue points) compared with mean correlation function (red points), computed using 10 mock realizations.
In addition to the \lya-SiIII line (see Fig.~\ref{fig:metalstack}), the effect 
of the SiII-SiIII shadow line is also visible on the 3D correlation for 
$\rperp \sim 0$ and $\rpar \sim 175 \hMpc$. 
Since bins for which $\rperp \sim 0$ contain small number of pairs compared to
bins at larger transverse separations, the presence of metal correlations on 
large scales do not bias BAO fitting procedures 
(Fig. 12 of \cite{delubac_baryon_2015}).

\section{Discussion \& Conclusions}
\label{sec:conc}

In this paper, we have presented the mock \lya~forest catalogs
for the Data-Release 11 of the SDSS-III. 
These mock spectra have been a fundamental tool for validating methods 
of the main BOSS \lya\  BAO measurements \cite{ 2013A&A...552A..96B, 2013JCAP...04..026S,2013JCAP...03..024K, delubac_baryon_2015}.

The analysis of the mocks presented in Section \ref{sec:covsec}
shows that they reproduce well the observed correlation function
of the transmission fluctuations as measured in \cite{delubac_baryon_2015}.
The measured covariance matrix of the mock correlation function is
comparable to that of the real data but does show
systematic deviations for both the variance and the off-diagonal
elements.  
As long as the errors are used in a consistent manner,
these differences do not prevent the mock spectra
from being useful to  verify that the analysis 
returns reasonable values for the statistical errors and to
show that there are no obvious biases in the measured
position of the BAO peak.

The deviations between mock and data covariances matrices
reflect the statistical properties
of individual mock spectra which, as shown in Section \ref{sec:tests},
differ from the real spectra. The differences are due
to the necessarily approximate treatment of the input power spectrum
at small scales. 
We have not simulated all the imperfections in the BOSS
pipeline, in particular the presence of Balmer artifacts.
We can anticipate, however, that
these artifacts will be removed with future improvements in the
pipeline.

Of more fundamental concern is the uncertain nature of the correct way
of introducing high column density systems and metallic absorbers.
Improvements in these aspects of the simulations will require further
study.

\begin{acknowledgments}

This research used resources of the National Energy Research Scientific 
Computing Center (NERSC), which is supported by the Office of Science of 
the U.S. Department of Energy under Contract No. DE-AC02-05CH11231.

The authors acknowledge the support of France Grilles for providing 
computing resources on the French National Grid Infrastructure.

This project was supported by the Agence Nationale de la Recherche
 under contract ANR-08-BLAN-0222. 
 The research leading to these results has 
 received funding from the European Union Seventh Framework Programme 
 (FP7/2007-2013) under grant agreement n¡ [PIIF-GA-2011-301665].

This work has been carried out thanks to the support of the A*MIDEX project (ANR-11-IDEX-0001-02) funded by the ``Investissements d'Avenir'' French Government program, managed by the French National Research Agency (ANR).

Funding for SDSS-III has been provided by the Alfred P. Sloan
Foundation, the Participating Institutions, the National Science
Foundation, and the U.S. Department of Energy Office of Science.
The SDSS-III web site is http://www.sdss3.org/.

SDSS-III is managed by the Astrophysical Research Consortium for the
Participating Institutions of the SDSS-III Collaboration including the
University of Arizona,
the Brazilian Participation Group,
Brookhaven National Laboratory,
University of Cambridge,
Carnegie Mellon University,
University of Florida,
the French Participation Group,
the German Participation Group,
Harvard University,
the Instituto de Astrofisica de Canarias,
the Michigan State/Notre Dame/JINA Participation Group,
Johns Hopkins University,
Lawrence Berkeley National Laboratory,
Max Planck Institute for Astrophysics,
Max Planck Institute for Extraterrestrial Physics,
New Mexico State University,
New York University,
Ohio State University,
Pennsylvania State University,
University of Portsmouth,
Princeton University,
the Spanish Participation Group,
University of Tokyo,
University of Utah,
Vanderbilt University,
University of Virginia,
University of Washington,
and Yale University.

\end{acknowledgments}

\bibliography{cosmo,cosmo_preprints}
\bibliographystyle{JHEP}

\appendix

\section{Access and usage of mocks}
\label{app:access}

The DR11 \lyaf\ mock catalogs are available at the SDSS public website \url{http://www.sdss.org/dr12/algorithms/lyman-alpha-mocks/}.
We describe in this section the mock data-sets format and their usage. Instead of providing the full set of realizations containing spectra ready to use, i.e., realizations with continua, noise and instrumental effects, only the raw absorption fields are provided. The ``expansion'' process is performed locally by the \emph{LyAMockExpander} package, available in \url{http://www.sdss3.org/svn/repo/boss/LyAMockExpander/}. This procedure reduces the amount of data that needs to be transferred, and gives the user the possibility of including or not some systematic effects. 

\subsection{Raw mock data format}
\label{sec:raw_format}

The first step consists in downloading raw mock fiels, 
containing absorption fields and high column density  
system information. 
A given raw file for a given spectrum is named: 
\texttt{mockrawShort-PLATE-MJD-FIBER.fits}. It contains a 
certain number of realizations of the same line of sight.
 
This file is organized as follows:
\begin{itemize}
\item \textbf{HDU 0}: header propagated form the real 
spectrum file, with additional information concerning the 
mock production (cosmological model and \lyaf\ clustering parameters). 
Mock related keywords are:
\begin{center}
\begin{tabular}{ll}
M\_Z     & Mock redshift   \\                                
M\_RA    & Mock RA [degrees]   \\
M\_DEC   & Mock dec [degrees]    \\                          
M\_OMEGAM & Mock Matter fraction $\Omega_m$         \\                          
M\_OMEGAK & Mock Curvature parameter $\Omega_k$         \\                          
M\_W     & Mock Dark-energy equation of state $w = p/\rho$                   \\                        
M\_H     & Mock Hubble constant $h= H_0/(100~\kms\iMpc)$                   \\                      
M\_NS    & Mock Primordial scalar spectral index $n_s$                   \\                    
M\_BIAS  & Mock \lyaf\ bias w.r.t. linear dark-matter power-spectrum           \\                  
M\_BETA  & Mock \lyaf\ redshift distortion parameter $\beta$                  \\                
M\_ALPHA & Mock \lyaf\ redshift evolution parameter $\alpha$                   \\               
M\_NEVOL & Mock number of redshift evolution steps \\       
M\_DV    & Mock grid spacing in km/s at $z_{\rm fid} = 2.6$  \\            
NMOCKS  &  Number of mock realizations in this file        
\end{tabular}
\end{center}

\item \textbf{HDU 1 to NMOCKS}: binary tables containing the absorption field for each realization. 

    \begin{tabular}{ll}
        \texttt{f} & the transmittance $F$ of the \lyaf\ \\
        \texttt{fdla} & the same as above but containing HCD         profiles
    \end{tabular}

\item \textbf{HDU NMOCKS+1 to 2*NMOCKS}: binary tables containing information about HCDs of the realizations.

    \begin{tabular}{ll}
        \texttt{x\_dla} & the comoving position of the system \\
        \texttt{z\_dla} & the corresponding redshift of the system \\
        \texttt{col\_dla} & the column density of the absorber, in cm$^{-2}$.
    \end{tabular}
\end{itemize} 

For DR11, the 100 realizations of a given forest are divided in 10 different files, containing 10 realizations each. 

\subsection{MockExpander}
\label{sec:mockexpander}

This package transforms raw absorption fields into quasar spectra 
including astrophysical and instrumental effects characterized in BOSS 
data. 

The \emph{MockExpander} was developed and compiled in JAVA language, and it is a stand alone, open-source version. 
%The \texttt{usefulFiles} folder contains files useful for the expansion process. The \texttt{src} folder contains all  modifiable source codes. The \texttt{lib} contains JAVA libraries, and \texttt{classes} contains the compiled classes in binary format.
The user chooses expansion options in the \texttt{Run.sh} script. We summarize these options here:

\begin{itemize}
\item \textbf{Raw\_Mock\_Directory}: the folder containing the raw format files
\item \textbf{Data\_Directory}: the folder containing the data spectra, containing individual exposures, in \texttt{spec}\footnote{\url{http://data.sdss3.org/datamodel/files/BOSS\_SPECTRO\_REDUX/RUN2D/spectra/PLATE4/spec.html}} format.
\item \textbf{Output\_Directory}: the folder were the \emph{MockExpander} will write the output mock files.
\item \textbf{Package\_Seed}: sets the random number generator seed based on which set of realizations.
\item \textbf{Initial(Final)\_Realization }: the user might want to expand a sub-set of the available realizations, these keywords allow the user to choose them, using their position in the file (1 to NMOCKS).
\item \textbf{ACTION}: This option allows the user to choose between start the expansion again (``Rewrite'') overwriting previous files, or continue (``Continue'') from the last mock produced.
\item \textbf{COOKING}: Set if the expansion will include high column density systems and/or metal absorption.
\item \textbf{COLS}: the user can choose the output columns of the expanded mock data-sets. They are fully described in Table~\ref{table:mock_format}.  
\end{itemize}

Once options are chosen, the user executes the \texttt{Run.sh} script to start the expansion.

\subsection{Expanded mocks data format}
\label{sec:mock_format}

The final output is a per-object FITS file
in the same format as real BOSS data.
The guiding principle is that if an analysis code can use the real data,
it should also be able to use the mock data without requiring any changes
other than the input file names. 

Mock files also contain additional information used in their construction. Table~\ref{table:mock_format} describes the content of HDU1. HDUs 0 (observing
headers) and 2 (metadata) are simply copies of the corresponding HDUs from
the real data files.  
Mock files do not receive HDU 3 from
real data (emission line fits and individual exposures).  Instead, HDU 3
contains the location of the high column density systems (it is a copy
of corresponding raw format HDU).

%--- Data Format ---
\begin{table*}
\begin{center}
\begin{tabular}{lll}
\bf{Column}  & \bf{Comment} & \bf{Symbol}\\
\hline
\multicolumn{3}{l}{\em{In both mocks and real data:}} \\
flux    & coadded calibrated flux [10$^{-17}$ ergs/s/cm$^2$/\AA] & $f$ \\
loglam  & $\log_{10}$(wavelength [\AA])  	& $\log_{10}\lambda$	  \\
ivar    & inverse variance of flux  & $\tilde{\sigma}_{\tilde{f}}$  \\
and\_mask    & mask & - \\
or\_mask     & mask & - \\
wdisp   & wavelength dispersion in dloglam units & $R_p,~R_w$ \\
sky     & subtracted sky flux [10$^{-17}$ ergs/s/cm$^2$/\AA] & $f_{\rm sky}$ \\
model   & pipeline best model fit used for classification  & - \\
 		& and redshift & \\
\hline
\multicolumn{3}{l}{\em{Only in the mocks:}} \\
mock\_F      & transmitted flux fraction [0--1]  & $F$ \\
mock\_Fdla   & transmitted flux fraction with Damped \lya\ systems & $F$ \\
mock\_Fmet   & transmitted flux fraction with metal absorption & $F$ \\
mock\_Fdlamet& transmitted flux fraction with Damped \lya\ systems & $F$ \\
			 & and metal absorption &  \\
mock\_meanF  & Mean transmitted flux fraction & $\bar{F}$ \\
mock\_ivar   & True inverse variance used to add noise &  $\sigma_{\tilde{f}}$  \\
mock\_contpca    & the PCA-based continuum model & $C$  \\
noise\_miscalib  & amount by which reported noise is wrong. & $r$ \\ 
				 & $<1$ means noise is underestimated. &   \\
mock\_miscalib   & amount by which the flux is purposefully miscalibrated & $M$  \\
mock\_missky     & amount of extra sky added to simulated sky mis-subtraction & $\delta f_{\rm sky}$ \\
\end{tabular}
\end{center}
\caption{Available columns for the HDU 1 of the mock spectra files. The \emph{MockExpander} allows the user to choose which columns the output mock files will contain, by the \textbf{COLS} keyword in the \texttt{Run.sh} script.  We describe how we compute each of these columns in \S~\ref{sec:expanded}.}
\label{table:mock_format}
\end{table*}

The \emph{MockExpander} will automatically organize output files in a per-realization and a per-plate folder scheme as for real data.

%\subsection{Data Release 10}
%\af{Explain why we can only provide DR10 mocks, what is the difference, write
%some warnings and add a couple of plots.}

% actuallly, at this point I don't think we should mention DR10, and make public
% the DR11 mocks together with DR12

\end{document}